\documentclass[a4paper,12pt]{article}
\usepackage{amssymb,latexsym}
\usepackage{fullpage}
\usepackage[dvips]{graphicx}
\usepackage{amssymb}
\usepackage[utf8]{inputenc}
\usepackage[english]{babel}
\usepackage{amsfonts, color}
\usepackage{amssymb}
\usepackage{amsthm}
\usepackage{cancel, slashed}

\newcommand{\ph}{\phantom}
\newcommand{\ba}{\begin{eqnarray}}

\newcommand{\ea}{\end{eqnarray}}
\newcommand{\be}{\begin{equation}}

\newcommand{\ee}{\end{equation}}

\newcommand{\mb}{\mathbf}
\newcommand{\mc}{\mathcal}

\newcommand{\p}{\partial}

\newcommand{\rar}{\rightarrow}

\newcommand{\nn}{\nonumber}
\newcommand{\e}{\epsilon}
\newcommand{\dar}{\Downarrow}

\newcommand{\ol}{\overline}
\newcommand{\im}{\mathrm{Im}}
\newcommand{\re}{\mathrm{Re}}

\makeatletter \@addtoreset{equation}{section} \makeatother

\input epsf

\begin{document}

%\ifpdf\DeclareGraphicsExtensions{.pdf, .jpg, .tif} \else%
%\DeclareGraphicsExtensions{.eps, .jpg} \fi
%QTO{titlepage}{%
\begin{titlepage}

   \thispagestyle{empty}
   \begin{flushright}
       \hfill{CERN-PH-TH/2009-38}\\
       \hfill{UCLA/09/TEP/45}\\
       \hfill{SU-ITP-09/11}\\
       \hfill{DFPD-09-TH-07 }\\
   \end{flushright}%

   %\vspace{5pt}
   \begin{center}
       { \huge{\textbf{More on $\mathcal{N}=8$ Attractors}}}

       \vspace{15pt}

       {\large{{\bf Anna Ceresole$^{\star}$, \ Sergio Ferrara$%
^{\diamondsuit,\spadesuit,\sharp}$,\\ Alessandra
Gnecchi$^{\clubsuit}$ and \ Alessio Marrani$^{\heartsuit}$ }}}

       \vspace{20pt}

       {$\star$ \it INFN, Sezione di Torino, and \\
       Dipartimento di Fisica Teorica, Universit\`{a} di Torino,\\
       Via Pietro Giuria, 1, I-10125
Torino, Italy\\
       \texttt{ceresole@to.infn.it}}

       \vspace{5pt}

       {$\diamondsuit$ \it Theory Division - CERN,\\
       CH
1211, Geneva 23, Switzerland\\
       \texttt{sergio.ferrara@cern.ch}}

\vspace{5pt}

        {$\spadesuit$ \it INFN - LNF, \\
         Via
Enrico Fermi 40, I-00044 Frascati, Italy}

         \vspace{5pt}

         {$\sharp$ \it Department of Physics and Astronomy,\\
       University of California, Los Angeles, CA USA}

         \vspace{5pt}

       {$\clubsuit$ \it
Dipartimento di Fisica ``Galileo Galilei", and \\
       INFN, Sezione di
Padova,\\
       Via Marzolo 8, I-35131 Padova, Italy\\
       \texttt{alessandra.gnecchi@pd.infn.it}
       \vspace{5pt}

       {$\heartsuit$
\it Stanford Institute for Theoretical Physics,\\
       Department of
Physics, 382 Via Pueblo Mall, Varian Lab,\\
       Stanford University,
Stanford, CA 94305-4060, USA\\\texttt{marrani@lnf.infn.it}}

       %\vspace{10pt}
    \vspace{20pt}

       {ABSTRACT}
       }

\end{center}

   \vspace{15pt}%

We examine few simple extremal black hole configurations of $\mathcal{N}=8$, $d=4$ supergravity. We first elucidate the relation between the BPS
Reissner-N\"{o}rdstrom black hole and the non-BPS Kaluza-Klein dyonic black hole. Their classical entropy, given by the Bekenstein-Hawking
formula, can be reproduced via the attractor mechanism by suitable choices of symplectic frame. Then, we display the embedding of the
axion-dilaton black hole into $\mathcal{N}=8$ supergravity.

   %\vspace{150pt}

\end{titlepage}
%\newpage\tableofcontents%\newpage

\section{\label{Intro}Introduction}

It has been known for some time \cite{Ferrara:1997uz} that extremal
BPS black hole (BH) states coming from string and M theory
compactifications to four and five dimensions, preserving various fractions of the original $\mathcal{N%
}=8$ supersymmetry, can be invariantly classified in terms of orbits
of the fundamental representations of the exceptional groups $E_{7(7)}$ and $%
E_{6(6)}$. These are the duality groups of the low energy actions,
whose discrete subgroups appear as symmetries of the
non-perturbative spectrum of BPS states \cite{Hull:1994ys}. These
orbits, which have been further studied in (\cite{Ferrara:1997ci,
Ferrara:2006xx,Bellucci:2006xz}), correspond to well defined
categories of allowed entropies of extremal BHs in $d=5$
and in $d=4$, given in terms of the cubic $E_{6(6)}$ invariant $\mathcal{I}%
_{3}$ (\cite{Ferrara:1997uz,Ferrara:2006xx,Andrianopoli:1997hb}) and
the quartic $E_{7(7)}$ invariant $\mathcal{I}_{4}$
(\cite{Kallosh:1996uy,Andrianopoli:1997pn,Ferrara:2006em}). There
are three types of orbits depending on whether the BH background
preserves $1/2$, $1/4$ or $1/8$ of the original supersymmetry. Only
$1/8$ BPS states have non vanishing entropy and regular horizons,
while $1/4$ and $1/2$ BPS configurations lead to vanishing classical
entropy.

The $\mathcal{N}=8$ attractors have been explored in
\cite{Ferrara:2006em} by solving the criticality condition for the
suitable BH effective potential and extending the lore of
$\mathcal{N}=2$ special K\"{a}hler geometry \cite{AM3}.

In this note we focus on some specific simple configurations in $\mathcal{N}%
=8$, $d=4$ supergravity which capture some representatives of the
regular (sometimes, called \textit{``large''}), \textit{i.e.} with
non-vanishing classical entropy, extremal BPS and non-BPS BH charge
orbits of the theory. One is the Reissner-N\"{o}rdstrom (RN) dyonic
BH, with electric and magnetic charge $e$ and $m$ respectively, and
Bekenstein-Hawking entropy (in unit of Planck mass) \cite{GH}
\begin{equation}
S_{RN}=\pi \left( e^{2}+m^{2}\right) .  \label{S-RN}
\end{equation}
Another one is the Kaluza-Klein (KK) dyonic BH, with a KK monopole
charge $p$ and a KK momentum $q$, which is dual to a $D0-D6$ brane
configuration in Type II A supergravity. Its Bekenstein-Hawking
entropy reads
\begin{equation}
S_{KK}=\pi \left| pq\right| .  \label{S-KK}
\end{equation}
One more interesting example is the extremal axion-dilaton BH,  a subsector of pure ${\mc N}=4$
 supergravity in $d=4$ which was considered in the past in \cite{K3,Kallosh:1993yg}.

Our aim is to show how the entropies of these BHs can be obtained in
the context of $\mathcal{N}=8$, $d=4$ supergravity by exploiting the
attractor mechanism \cite{AM1,strominger2,AM3,FGK} for extremal BPS
and non BPS BHs . Earlier studies for some specific cases where
examined in \cite{KO1,KO2}.

It is in fact known that while the BH charge configuration with
entropy given by (\ref{S-RN}) is 1/8 BPS \cite{GH}, the entropy
(\ref{S-KK}) is related to a non BPS one. Indeed, the $E_{7\left(
7\right) }$ quartic invariant $\mathcal{I}_{4}$ on these
configurations reduces to
\begin{eqnarray}
\sqrt{{\mathcal{I}_{4}^{RN}}} &=&e^{2}+m^{2};  \label{I4-RN} \\
\sqrt{-\mathcal{I}_{4}^{KK}} &=&\left| pq\right| .  \label{I4-KK}
\end{eqnarray}
In particular we note that , if the magnetic (or electric) charge is
switched off, the RN BH remains regular, whereas the KK BH reaches
zero entropy ($\mathcal{I}_{4}=0$) and becomes 1/2 BPS
\cite{Ferrara:1997ci}.

The simplest way to obtain these configurations is to observe that
the BPS and non-BPS charge orbits with $\mathcal{I}_{4}\neq 0$
%supporting non degenerate, asymptotically flat, spherically symmetric, static dyonic extremal BHs
in $\mathcal{N}=8$, $d=4$ supergravity are given by
\cite{Ferrara:1997uz}
\begin{eqnarray}
{\mathcal{O}}_{1/8-BPS} &:&\frac{E_{7\left( 7\right) }}{E_{6\left(
2\right)
}},~\quad \quad \mathcal{I}_{4}>0;  \label{O-1/8-BPS-large} \\
{\mathcal{O}}_{non-BPS} &:&\frac{E_{7\left( 7\right) }}{E_{6\left(
6\right) }},~\quad \quad \mathcal{I}_{4}<0.  \label{O-non-BPS}
\end{eqnarray}
The moduli spaces corresponding to the above disjoint orbits  are
\cite{Ferrara-Marrani-1} \ba
{\mc M}_{1/8-BPS}&=&\frac{E_{6(2)}}{SU(6)\times SU(2)}\nn\\
{\mc M}_{non-BPS}&=&\frac{E_{6(6)}}{USp(8)}\, . \ea

Hence, a convenient representative of these orbits is given by the (unique) $%
E_{6}$-singlets in the decomposition of the fundamental representation $%
\mathbf{56}$ of $E_{7\left( 7\right) }$ into the two relevant
non-compact real forms of $E_{6}$: %\begin{eqnarray}
%\left( \frac{1}{8}\text{\textit{-}}\right) \text{\textit{BPS}}\mathit{~}%
%\text{\textit{``large''}} &:&\left\{
%\begin{array}{l}
%E_{7\left( 7\right) }\rightarrow E_{6\left( 2\right) }\times U\left(
%1\right) ; \\
%\\
%\mathbf{56}\rightarrow \left( \mathbf{27},1\right) +\left( \overline{\mathbf{%
%27}},-1\right) +\left( \mathbf{1},3\right) +\left(
%\mathbf{1},-3\right) ;
%\end{array}
%\right.  \label{BPS-branching} \\
%&&  \notag \\
%\text{\textit{non-BPS}} &:&\left\{
%\begin{array}{l}
%E_{7\left( 7\right) }\rightarrow E_{6\left( 6\right) }\times
%SO\left(
%1,1\right) ; \\
%\\
%\mathbf{56}\rightarrow \left( \mathbf{27},1\right) +\left( \mathbf{27}%
%^{\prime },-1\right) +\left( \mathbf{1},3\right) +\left( \mathbf{1}%
%,-3\right) ,
%\end{array}
%\right.  \label{non-BPS-branching}
%\end{eqnarray}
\begin{eqnarray}
RN\quad\quad{\mathcal{O}}_{1/8-BPS} &:&\left\{
\begin{array}{l}
E_{7\left( 7\right) }\rightarrow E_{6\left( 2\right) }\times U\left(
1\right) ; \\
\\
\mathbf{56}\rightarrow \left( \mathbf{27},1\right) +\left( \mathbf{1}%
,3\right) +\left( \overline{\mathbf{27}},-1\right) +\left( \mathbf{1}%
,-3\right) ;
\end{array}
\right.   \label{BPS-branching} \\
&&~  \nn \\
KK\quad\quad{\mathcal{O}}_{non-BPS} &:&\left\{
\begin{array}{l}
E_{7\left( 7\right) }\rightarrow E_{6\left( 6\right) }\times
SO\left(
1,1\right) ; \\
\\
\mathbf{56}\rightarrow \left( \mathbf{27},1\right) +\left( \mathbf{1}%
,3\right) +\left( \mathbf{27}^{\prime },-1\right) +\left( \mathbf{1^\prime }%
,-3\right) ,
\end{array}
\right.   \label{non-BPS-branching}
\end{eqnarray}
where the $U\left( 1\right) $ charges and $SO\left( 1,1\right) $
weights are indicated, and the prime denotes the contravariant
representations. Notice that, consistently with the group factors
$U\left( 1\right) $ and $SO\left( 1,1\right) $, $\mathbf{27}$ is
complex for $E_{6\left( 2\right) }$, whereas
it is real for $E_{6\left( 6\right) }$ . Both $E_{6(2)}\times U(1)$ and $%
E_{6\left( 6\right) }\times SO\left( 1,1\right) $ are maximal
non-compact subgroups of $E_{7\left( 7\right) }$, with symmetric
embedding.\medskip\

Our result is simply stated as follows.

The two extremal BH charge configurations determining the embedding
of RN and KK extremal BHs into $\mathcal{N}=8$, $d=4$ supergravity
with entropies (\ref{S-RN}) and (\ref{S-KK}), are given by the two $E_{6}$%
-singlets in the decompositions (\ref{BPS-branching}) and (\ref {non-BPS-branching}).

The two situations can be efficiently associated to two different parametrizations
of the real symmetric scalar manifold $\frac{%
E_{7\left( 7\right) }}{SU\left( 8\right) }$ ($dim_{\mathbb{R}}=70$, \textit{%
rank}$=7$) of $\mathcal{N}=8$, $d=4$ supergravity.

For the branching (\ref{BPS-branching}), pertaining to the RN
extremal BH, the relevant parametrization
is the $SU\left( 8\right) $%
-covariant one. This corresponds to the Cartan's decomposition
basis, where the coset coordinates $\phi _{ijkl}$ ($i=1,...8$) sit
in the four-fold antisymmetric self-real irrep $\mathbf{70}$ of
$SU(8)$. The attractor
mechanism implies that at the horizon
\begin{equation}
\phi _{ijkl,H}=0,  \label{RN-AM}
\end{equation}
\textit{i.e.} the scalar configuration at the event horizon of the
1/8-BPS extremal BH is given by the origin of $\frac{E_{7\left(
7\right) }}{SU\left( 8\right) }$. Some care should be taken with regards to
\textit{``flat''} directions \cite
{Andrianopoli:1997pn,Ferrara-Marrani-1}. Due to the existence of the
moduli space $\frac{E_{6(2)}}{SU(6)\times SU(2)}$ ($dim_{\mathbb{R}}=40$, \textit{%
rank}$=4$) of the $\frac{1}{8}$-BPS attractor solutions, strictly speaking $%
40$ scalar degrees of freedom out of $70$ are actually undetermined
at the event horizon of the given $\frac{1}{8}-$BPS RN extremal BH.
In other words, $40$ real scalar degrees of freedom, spanning the
quaternionic symmetric coset $\frac{E_{6(2)}}{SU(6)\times SU(2)}$
(which is the $c$\textit{-map}
\cite{CFG1} of the vector multiplets' scalar manifold of $\mathcal{N}=2$, $%
d=4$ \textit{``magic''} supergravity based on $J_{3}^{\mathbb{C}}$),
can be set to any real value, without affecting the RN BH entropy
(\ref{S-RN}).

It should be noticed that, consistently with the Gaillard-Zumino
formulation of electric-magnetic duality in presence of
scalar fields \cite{GZ1}, the solution (\ref {RN-AM}) to the
attractor equations is the only one allowed in presence of a
\textit{compact} underlying symmetry (in this case $U\left( 1\right)
$).

On the other hand, the best parametrization for the branching (\ref{non-BPS-branching}), pertaining to
the KK extremal BH, is given by by the KK radius
\begin{equation}
r_{KK}\equiv \mathcal{V}^{1/3}\equiv e^{2\varphi },  \label{r-KK}
\end{equation}
by the $42$ real scalars $\psi _{ijkl}$\ ($i=1,...8$) sitting in the $%
\mathbf{42}$ of $USp\left( 8\right) $, and by the $27$ real \textit{%
axions} $a^{I}$ ($I=1,\ldots ,27$) sitting in the $\mathbf{27}$ of $%
USp\left( 8\right) $ (or equivalently, in the $\mathbf{27}$ of
$E_{6(6)}$).

In virtue of the attractor mechanism, the KK radius is stabilized as
follows \cite{Ceresole:2007rq}:
\begin{equation}
r_{KK,H}^{3}\equiv \mathcal{V}_{H}\equiv e^{6\varphi _{H}}=4\left| \frac{q}{p%
}\right| ,  \label{r^3_H-KK}
\end{equation}
while all \textit{axions }vanish:
\begin{equation}
a_{H}^{I}=0.  \label{a_H-KK}
\end{equation}
The $42$ real scalars $\psi _{ijkl}$ are actually undetermined at
the event horizon of the non-BPS KK BH, without affecting its
entropy (\ref
{S-KK}). Indeed, they span the moduli space $\frac{E_{6(6)}}{USp(8)}$ ($dim_
{%
\mathbb{R}}=42$, \textit{rank}$=6$) of the non-BPS attractor solutions, which is the real symmetric scalar manifold of
$\mathcal{N}=8$, $d=5$ supergravity \cite{Ferrara-Marrani-1}.

It should be clear from our discussion that the possibility of having
a non-vanishing scalar stabilized at the horizon of the KK extremal
BH is related to the presence of a singlet in the relevant
decomposition of the $70$ scalars. This in turn is related to the
existence of an underlying non-compact symmetry ($SO\left(
1,1\right) $ in the present case), admitting no compact
sub-symmetry.

An alternative way to obtain eqs. (\ref{S-RN}) and (\ref{S-KK})  is to use
appropriate truncations  for the  bare charges in the corresponding expression of
the quartic invariant $\mathcal{I}_{4}$, which is known to be related to the
Bekenstein-Hawking entropy by the formula
\begin{equation}
S=\sqrt{\left| \mathcal{I}_{4}\right| }.
\end{equation}

The manifestly $SU(8)$-invariant expression of $\mathcal{I}_{4}$
reads as follows:
\begin{equation}
\mathcal{I}_{4}=Tr \left( ZZ^{\dag }\right) ^{2} -\frac{1}{4}%
Tr^{2}\left( ZZ^{\dag }\right) +8Re  Pf\left( Z\right) \label{US?},
\end{equation}
where $Z\equiv Z_{AB}\left( \phi \right) $ is the central charge
$8\times 8$ skew-symmetric matrix. Since (\ref{US?}) is
moduli-independent, it can be
evaluated at $\phi =0$ without loss of generality, and in such a case $%
Z_{AB}$ is replaced by $Q_{AB}$, the \textit{bare} charge matrix in the $%
SU\left( 8\right) $ basis.

Considering the RN black hole, we will see that a suitable truncation of the $\mc N=8$ \textit{bare} charge matrix $Q_{AB}$ ($A,B=1,\ldots8)$,
reduces it to the form

\begin{equation}
Q_{AB}^{RN}\rightarrow \left( z\epsilon _{ab},0\right) ,~z\equiv
e+im,
\end{equation}
where $a,b=1,2$ and $\epsilon
^{T}=-\epsilon $). Thus one obtains
\begin{equation}
\mathcal{I}_{4}=\left| z\right| ^{4}=\left( e^{2}+m^{2}\right) ^{2},
\end{equation}
which is nothing but Eq. (\ref{I4-RN}) and it is also the same
result  as in \textit{pure} $\mathcal{N}=2$, $d=4$ supergravity,
which has a $U(1)$ global $\mathcal{R}$-symmetry \cite{GH}.

On the other hand, the manifestly $E_{6\left( 6\right) }$-invariant
expression of $\mathcal{I}_{4}$ in terms of the cubic invariant ${\mc I}_3$, as function of the bare electric and magnetic charges  is given by
\cite{Ferrara:1997uz,Ferrara:2006yb,Bellucci:2006xz}:
\begin{equation}
\mathcal{I}_{4}=-\left( p^{0}q_{0}+p^{i}q_{i}\right) ^{2}+4\left[ q_{0}%
\mathcal{I}_{3}\left( p\right) -p^{0}\mathcal{I}_{3}\left( q\right)
+\left\{
\mathcal{I}_{3}\left( p\right) ,\mathcal{I}_{3}\left( q\right) \right\} %
\right] .
\end{equation}
By truncating the fluxes in such a way that
\begin{equation}
p^{i}=0=q_{i},
\end{equation}
one obtains ($p^{0}\equiv p$, $q_{0}\equiv q$)
\begin{equation}
\mathcal{I}_{4}=-\left( pq\right) ^{2},
\end{equation}
which now coincides with Eq. (\ref{I4-KK}).

We will show that there is yet another way to obtain the two entropies for RN and KK black holes
(\ref{S-RN}) and (\ref{S-KK}) . This consists in using the attractor equations for the effective black hole potential
$\frac{\partial V_{BH}}{\partial \phi}=0 $ and the expression of the entropy as the value of such potential  at the critical point,
\begin{equation}
\label{entropia}
S=\pi \left. V_{BH}\right| _{crit}.
\end{equation}
\medskip\

The plan of this paper is as follows.

In Sect. \ref{Sympl-Frames} we consider various bases of $\mathcal{N}=8$, $%
d=4$ supergravity, namely the $SL\left( 8,\mathbb{R}\right) $,
$SU\left( 8\right) $- and $USp(8)$-covariant ones, exploiting the
relevant branchings of the $U$-duality group $E_{7\left( 7\right)
}$. Then, Sect. \ref{SL(8,R)} is devoted to the computation of the
fundamental quantities for the geometry of the scalar manifold
$\frac{E_{7\left( 7\right) }}{SU\left( 8\right) }$ in the $SL\left(
8,\mathbb{R}\right) $-covariant basis.
Then, Sect. \ref{jazz} analyses  the $E_{6\left(6\right) }$-covariant basis,  with the goal of exhibiting the connection with $\mc N=8$,  $%
d=5$ supergravity: the $d=4$ effective BH potential is recast in a manifestly $d=5$ covariant
form. Moreover,  the charge configurations of this potential leading to vanishing  axion fields  are studied
along with the corresponding attractor solutions.
In Sect. \ref{Axion-Dilaton-d=4} the embedding of the axion-dilaton extremal BH in $%
\mathcal{N}=8$, $d=4$ supergravity, through an intermediate embedding into $%
\mathcal{N}=4$, $d=4$ theory with $6$ vector multiplets, is
analyzed. Finally, Sect. \ref{Conclusion} contains an outlook, as
well as some concluding comments and remarks. The paper also contains in an Appendix the
embedding of the $d=5$ uplift of the $stu$ model (the
so-called $\left( SO\left( 1,1\right) \right) ^{2}$ model) into
$d=5$ maximal supergravity.

\section{\label{Sympl-Frames}Symplectic Frames}

The de Wit-Nicolai \cite{de Wit:1982ig} formulation of
$\mathcal{N}=8$, $d=4$ supergravity is based on a symplectic frame
where the maximal non-compact symmetry of the Lagrangian is
$SL\left( 8,\mathbb{R}\right) $ \cite{HW}, according to the
decomposition
\begin{equation}
\begin{array}{l}
E_{7\left( 7\right) }\rightarrow SL\left( 8,\mathbb{R}\right) , \\
\\
\mathbf{56}\rightarrow \mathbf{28}+\mathbf{28}^{\prime },
\end{array}
\label{E_7(7)-->SL(8,R)}
\end{equation}
where $SL\left( 8,\mathbb{R}\right) $ is a maximal non-compact
subgroup
%(with symmetric embedding)
of $E_{7\left( 7\right) }$, and $\mathbf{28}$ is its two-fold
antisymmetric irreducible representation. Since the theory is \textit{pure}, the $\mathcal{R}$%
-symmetry, namely $SU\left( 8\right) $, is the stabilizer of the
scalar manifold. It is not a symmetry of the Lagrangian, but only of
the equations of motion. The maximal compact symmetry of the Lagrangian
is the intersection of $SL\left( 8,\mathbb{R}\right) $ with
$SU\left( 8\right) $, which is  $SO\left( 8\right) $ (the maximal compact subgroup
of $SL\left( 8,\mathbb{R}\right) $ itself).

Another symplectic frame corresponds to the decomposition (\ref
{non-BPS-branching}). In this case, the maximal non-compact symmetry
of the Lagrangian is $E_{6\left( 6\right) }\times SO\left(
1,1\right) \otimes _{s}T_{27}$, with ``$\otimes _{s}$''
denoting the semi-direct group product and $T_{27}$ standing for the
$27$-dimensional Abelian subgroup of $E_{7\left( 7\right) }$. The maximal compact symmetry is now
$USp\left( 8\right) $, which is also  the maximal compact symmetry of the Lagrangian.
Note that all terms in the Lagrangian are $SU(8)$ invariant, with
the exception of the vector kinetic terms, which are $SU\left( 8\right) $%
-invariant only on-shell.

%Let us further decompose
%\%begin{equation}
%SL\left( 8,\mathbb{R}\right) \rightarrow SL\left(
%6,\mathbb{R}\right) \times SL\left( 2,\mathbb{R}\right) \times
%SO\left( 1,1\right) \,,
%\end{equation}
%so that
%\begin{equation}
%\begin{array}{l}
%\mathbf{28}\rightarrow \left( \mathbf{15},\mathbf{1},1\right)
%+\left(
%\mathbf{6},\mathbf{2},-1\right) +\left( \mathbf{1},\mathbf{1},-3\right) , \\
%\\
%\mathbf{28}^{\prime }\rightarrow \left( \mathbf{15}^{\prime },\mathbf{1}%
%,-1\right) +\left( \mathbf{6}^{\prime },\mathbf{2},1\right) +\left( \mathbf{1%
%},\mathbf{1},3\right) ,
%\end{array}
%\label{SL(8,R)-->SL(6,R)xSL(2,R)xSO(1,1)}
%\end{equation}
%where $SL\left( 6,\mathbb{R}\right) \times SL\left(
%2,\mathbb{R}\right) \times SO\left( 1,1\right) $ is a maximal
%non-compact subgroup (with symmetric embedding) of $SL\left(
%8,\mathbb{R}\right) $.

Let us decompose  $E_{7(7)}$ along  two different  maximal non-compact subgroups according to  the following diagram:
\be
\label{diagram1}
\begin{array}{ccc}
E_{7( 7)}&\longrightarrow &SL( 8,\mathbb{R})\\
          &    & \\
\downarrow&                 &\downarrow \\
& & \\
E_{6(6)}\times SO(1,1)\quad\quad&\longrightarrow& \quad\quad SL( 6,\mathbb{R}) \times SL(2,\mathbb{R}) \times SO( 1,1)\,  .
\end{array}
\ee
If one goes first horizontally, the $\mathbf 56$ of $E_{7(7)}$  decomposes as
\begin{equation}
\mathbf{56}\rightarrow \mathbf{28}+\mathbf{28}^{\prime }\rightarrow
\left\{
\begin{array}{l}
\left( \mathbf{15},\mathbf{1},1\right) +\left( \mathbf{6},\mathbf{2}%
,-1\right) +\left( \mathbf{1},\mathbf{1},-3\right) + \\
\\
+\left( \mathbf{15}^{\prime },\mathbf{1},-1\right) +\left( \mathbf{6}%
^{\prime },\mathbf{2},1\right) +\left(
\mathbf{1},\mathbf{1},3\right) .
\end{array}
\right.
\label{E_7(7)-->SL(8,R)-further}
\end{equation}
Alternatively, one can first go downward, and use that
\begin{equation}
\begin{array}{l}
E_{6(6) }\rightarrow SL( 6,\mathbb{R}) \times
SL\left( 2,\mathbb{R}\right) ; \\
\\
\mathbf{27}\rightarrow \left( \mathbf{15},\mathbf{1}\right) +\left( \mathbf{6%
}^{\prime },\mathbf{2}\right) , \\
\\
\mathbf{1}\rightarrow \left( \mathbf{1},\mathbf{1}\right) ,
\end{array}
\label{E_6(6)-->SL(6,R)xSL(2,R)}
\end{equation}
thus obtaining:
\begin{equation}
\mathbf{56}\rightarrow \left( \mathbf{27},1\right) +\left( \mathbf{1}
,3\right) +\left( \mathbf{27}^{\prime },-1\right) +\left( \mathbf{1}
,-3\right) \rightarrow \left\{
\begin{array}{l}
\left( \mathbf{15},\mathbf{1},1\right) +\left( \mathbf{6}^{\prime },\mathbf{2
},1\right) +\left( \mathbf{1},\mathbf{1},3\right) + \\
\\
+\left( \mathbf{15}^{\prime },\mathbf{1},-1\right) +\left( \mathbf{6},%
\mathbf{2},-1\right) +\left( \mathbf{1},\mathbf{1},-3\right) .
\end{array}
\right.
\label{E_7(7)-->E_6(6)xSO(1,1)-further}
\end{equation}
Therefore, either way on the diagram and irrespectively of the intermediate decomposition, one obtains the same
irreducible representations of $SL\left( 6,\mathbb{R}\right) \times SL\left( 2,\mathbb{R}%
\right) \times SO\left( 1,1\right) $, which enjoyes a unique embedding in the $U$-duality group
$E_{7\left(7\right) }$.
In particular, one sees that the singlets are indeed the same in the two cases, and the alternative
decompositions are related by the interchange of
$\left( \mathbf{15},\mathbf{1},1\right) $ with $\left( \mathbf{15}^{\prime },%
\mathbf{1},-1\right) $.
Then one concludes that these two formulations, corresponding to two different symplectic frames,  can be
interchanged by dualizing $15$ out of the $28$ vector fields.\medskip

An analogous argument holds if one decomposes $E_{7(7)}$ according two two different maximal compact subgroups along the diagram
\be
\label{diagram2}
\begin{array}{ccc}
E_{7( 7)}&\longrightarrow &SU( 8)\\
                &                          & \\
\downarrow&                 &\downarrow \\
& & \\
E_{6(2)}\times U(1)\quad\quad&\longrightarrow& \quad\quad SU( 6) \times SU(2) \times U(1)\,  .
\end{array}
\ee
This time, going first horizontally along the diagram, the result reads:
\begin{equation}
\mathbf{56}\rightarrow \mathbf{28}+\overline{\mathbf{28}}\rightarrow
\left\{
\begin{array}{l}
\left( \mathbf{15},\mathbf{1},1\right) +\left( \mathbf{6},\mathbf{2}%
,-1\right) +\left( \mathbf{1},\mathbf{1},-3\right) + \\
\\
+\left( \overline{\mathbf{15}},\mathbf{1},-1\right) +\left( \overline{%
\mathbf{6}},\mathbf{2},1\right) +\left(
\mathbf{1},\mathbf{1},3\right) .
\end{array}
\right.
\label{E_7(7)-->SU(8)-further}
\end{equation}

Equivalently, one can first go vertically on the diagram and use
\begin{equation}
\begin{array}{l}
E_{6\left( 2\right) }\rightarrow SU\left( 6\right) \times SU\left(
2\right) ;
\\
\\
\mathbf{27}\rightarrow \left( \mathbf{15},\mathbf{1}\right) +\left(
\overline{\mathbf{6}},\mathbf{2}\right) , \\
\\
\mathbf{1}\rightarrow \left( \mathbf{1},\mathbf{1}\right) ,
\end{array}
\label{E_6(2)-->SU(6)xSU(2)}
\end{equation}
thus obtaining:
\begin{equation}
\mathbf{56}\rightarrow \left( \mathbf{27},1\right) +\left( \overline{\mathbf{%
27}},-1\right) +\left( \mathbf{1},3\right) +\left(
\mathbf{1},-3\right) \rightarrow \left\{
\begin{array}{l}
\left( \mathbf{15},\mathbf{1},1\right) +\left( \overline{\mathbf{6}},\mathbf{%
2},1\right) +\left( \mathbf{1},\mathbf{1},3\right) + \\
\\
+\left( \overline{\mathbf{15}},\mathbf{1},-1\right) +\left( \mathbf{6},%
\mathbf{2},-1\right) +\left( \mathbf{1},\mathbf{1},-3\right) .
\end{array}
\right.
\label{E_7(7)-->E_6(2)xU(1)-further}
\end{equation}
Again, either of the two alternative branchings in  (\ref{diagram2}) , which  are  related by the interchange of
$\left( \mathbf{15},\mathbf{1},1\right) $ with $\left( \overline{\mathbf{15}}%
,\mathbf{1},-1\right) $, yield the same decomposition
into irreducible representations of $SU\left( 6\right) \times SU\left( 2\right) \times
U(1)$. Moreover, the $U(1)$ singlet which commutes with $SU\left( 6\right) \times SU\left( 2\right) $ is the same as the
one which commute with $E_{6\left( 2\right) }$.
\medskip

Let us now turn to the scalar sector.
As mentioned above, the coordinate system for  the scalar manifold
$\frac{E_{7\left( 7\right) }}{SU(8)}$ based on the Cartan decomposition, the real  scalars $\phi_{ijkl}$ sit in the $\mathbf{70}$ (
four-fold antisymmetric and self-real irreducible representation) of $SU(8)$ with $i=1,\ldots, 8$. The
embedding of the RN extremal BH is related to the further decomposition
\begin{equation}
\begin{array}{l}
SU(8)\rightarrow SU\left( 6\right) \times SU\left( 2\right) \times U(1), \\
\\
\mathbf{70}\rightarrow \left( \mathbf{20},\mathbf{2},0\right)
+\left(
\mathbf{15},\mathbf{1},-2\right) +\left( \overline{\mathbf{15}},\mathbf{1}%
,2\right) .
\end{array}
\label{SU(8)-->SU(6)xSU(2)xU(1)-70}
\end{equation}
On the other hand, for describing the KK extremal BH one decomposes $SU(8)$ under its maximal subgroup $USp(8)$:
\begin{equation}
\begin{array}{l}
SU(8)\rightarrow USp(8), \\
\\
\mathbf{70}\rightarrow \mathbf{42}+\mathbf{27}+\mathbf{1},
\end{array}
\label{SU(8)-->USp(8)-70}
\end{equation}
where $\mathbf{42}$ and $\mathbf{27}$ are
respectively the four-fold and two-fold antisymmetric irreducible representations
(both skew-traceless and self-real) of $USp\left( 8\right) $.

The crucial difference between (\ref
{SU(8)-->SU(6)xSU(2)xU(1)-70}) and (\ref{SU(8)-->USp(8)-70}) is that
the latter decomposition contains a real singlet, whereas the first
one does not.  This is related to an underlying maximal compact  ($U\left(
1\right) $  symmetry which is present for (\ref {SU(8)-->SU(6)xSU(2)xU(1)-70}) and not for
(\ref{SU(8)-->USp(8)-70}).  This feature explains  the different behaviour of the two solutions at the attractor point: the RN solution has
the behaviour (\ref{RN-AM}) while the KK solution is given by
(\ref{r^3_H-KK})-(\ref{a_H-KK}).

\section{\label{SL(8,R)}$SL\left( 8,\mathbb{R}\right)$-Basis}
%\subsection{Symplectic sections.}
In this section we aim at making contact between the symplectic
formalism for extended supergravities reviewed in \cite
{Andrianopoli:2006ub} and the original formulation of $\mathcal N=8$
supergravity of \cite{de Wit:1982ig} for some of the key geometrical
objects that are relevant for the present investigation (see also
\cite{KS-N=8} for recent developments).

We start by considering the coset representative for $E_{7(7)/SU(8)}$, which  is parametrized as \cite{de Wit:1982ig}
\ba\label{} \mc V&=& \left(
\begin{array}{cc}
u_{ij}^{IJ}&v_{ijKL}\\
v^{klIJ}&u^{kl}_{KL}\, .
\end{array}
\right) \ea The sub-matrices $u$ and $v$ carry indices of  both
$E_{7(7)}$ and $SU(8)$ ($I=1,\ldots,8$, $I=1,\ldots,8$)
%\ba\label{} u_{ij}^{IJ}\ ,\quad v^{ijIJ}\ ,\ea
but one can choose a suitable $SU(8)$ gauge for the fields, and
then retain only manifest invariance with respect to the rigid
diagonal subgroup of $E_{7(7)}\times SU(8)$, without distinction
among the two types of indices. Comparing the notation of \cite{de Wit:1982ig} (in
particular the appendix B) with the symplectic formalism of \cite
{GZ1,Andrianopoli:2006ub}, we can identify
\ba\label{} \left\{
\begin{array}{c}
\phi_{0}\equiv u\nn\\
\phi_{1}\equiv v
\end{array}
\right.\qquad\rar\qquad\left.
\begin{array}{l}
u_{ij}^{\ph{ij}kl}=(P^{-1/2})_{ij}^{\ph{ij}kl}\ ,\\
v^{ijkl}=-(\bar P^{-1/2})^{ij}_{\ph{ij}mn}\bar y^{mnkl}
\end{array}\right.
\ea so that \ba \left\{
\begin{array}{c}
\mb f=\frac1{\sqrt2}(\phi_{0}+\phi_{1})=\frac1{\sqrt2}(u+v)\\
i\mb h=\frac1{\sqrt2}(\phi_{0}-\phi_{1})=\frac1{\sqrt2}(u-v)
\end{array}
\right.\ . \ea Since sections are sub-matrices of the symplectic
representation, relatively to electric and magnetic subgroups, their
explicit indices components are given by \ba\label{sections}
f_{ij}^{\ph{ij}kl}=\frac1{\sqrt2}\left(
(P^{-1/2})^{\ph{ij}kl}_{ij}-(\bar P^{-1/2})^{ij}_{\ph{ij}mn}\bar
y^{mnkl}
\right)\ ,\nn\\
h_{ij,kl}=\frac{-i}{\sqrt2}\left( (P^{-1/2})^{\ph{ij}kl}_{ij}+(\bar
P^{-1/2})^{ij}_{\ph{ij}mn}\bar y^{mnkl} \right)\ , \ea
where, in matrix notation,
\ba\label{} P=1-YY^{\dag}\
,\qquad Y=B\frac{\tanh \sqrt{B^{\dag}B}}{\sqrt{B^{\dag}B}}\ , \qquad
B_{ij,kl}=-\frac{1}{2\sqrt2}\phi_{ijkl}\ , \ea the last definition
coming from the choice of the symmetric gauge for the coset
representative in Eq. (B.1) of \cite {de Wit:1982ig}. If one defines
\ba\label{} \tilde P=1-Y^{\dag}Y\ , \ea and uses
the identity
\ba\label{} (\tilde
P^{-1/2})Y^{\dag}=Y^{\dag}(P^{-1/2})\, , \ea
the following simple expressions for $\mathbf{f}$ and $\mathbf{h}$ are finally
achieved:
\ba\label{sezione f}
\mb f&=&\frac1{\sqrt2}\left[ P^{-1/2}-(\tilde P^{-1/2})Y^{\dag} \right]=\frac1{\sqrt2}[1-Y^{\dag}]\frac1{\sqrt{1-YY^{\dag}}}\ ,\\
\label{sezione h}
\mb h&=&-\frac{i}{\sqrt2}\left[P^{-1/2}+(\tilde
P^{-1/2})Y^{\dag}\right]=-\frac{i}{\sqrt2}[1+Y^{\dag}]\frac1{\sqrt{1-YY^{\dag}}}\ . \ea
The above notations  are such that \ba\label{}
P^{1/2}&=&\sqrt{1-YY^{\dag}}\qquad\rar\qquad  P_{ij}^{\ph{ij}kl}=\delta_{ij}^{kl}-y_{ijmn}\bar y^{mnkl}\nn\\
\tilde P^{1/2}&=&\sqrt{1-Y^{\dag}Y}\qquad\rar\qquad \bar
P^{kl}_{\ph{kl}ij}=\delta^{kl}_{ij}-\bar y^{klmn}y_{mnij} \ea

It is easily checked that the  symplectic sections satisfy the usual relations \ba\label{}
i(\mb{f^{\dag}h-h^{\dag}f})&=&1\ ,\nn\\
\mb{h}^{T}\mb f-\mb f^{T}\mb h&=&0\ . \ea These are obtained writing
the symplectic sections as in (\ref{sezione f}) and (\ref{sezione
h}), and using the identity \ba Y\tilde{P}^{-1}=P^{-1}Y\ . \ea

The kinetic matrix is given in terms of the symplectic sections by
\cite{Andrianopoli:2006ub} \ba\label{} \mc N&=&\mb h\mb f^{-1}\ .\ea
Therefore, Eqs. (\ref{sezione f}) and (\ref{sezione h}) yield
\ba\label{}
\mc N&=&-i\,[1+Y^{\dag}]\frac1{\sqrt{1-YY^{\dag}}} \sqrt{1-YY^{\dag}}\frac1{1-Y^{\dag}}=\nn\\
&=&-i\,\frac{1+Y^{\dag}}{1-Y^{\dag}}\nn\\
&\dar&\nn\\
\mc N_{ij|kl}&=&-i(\delta_{mn}^{kl}+\bar y^{mnkl})(\delta_{ij}^{mn}-\bar y^{ijmn})^{-1}\ .\
\ea
%We remind that the kinetic matrix can be written as, once the symplectic section have been normalized as above,
%\ba\label{} \mc
%N_{\Lambda\Sigma}&=&i(\phi_{0}^{\dag}+\phi_{1}^{\dag})^{-1}(\phi_{0}^{\dag}-\phi_{1}^{\dag})\
%, \ea and from the definitions above we get to \ba\label{}
%\mc N_{\Lambda\Sigma}&=&i(u^{\dag}+v^{\dag})^{-1}(u^{\dag}-v^{\dag})=\nn\\
%&=&\left( P^{-1/2} -Y(\tilde P^{-1/2})\right)^{-1}\left( P^{-1/2}+Y(\tilde P)^{-1/2} \right)=\nn\\
%&=&\frac{1+Y}{1-Y}\ ,\nn\\
%&\dar&\nn\\
%\mc
%N_{ij|kl}&=&i(\delta_{ij}^{mn}-y_{ijmn})^{-1}(\delta_{mn}^{kl}+y_{mnkl})\
%. \ea

%\subsection{Central charge function.}
We now turn to the central charge function, which is defined by
\ba\label{}
Z_{ij}&=&f^{\ph{ij}kl}_{ij}q_{kl}-h_{ij|kl}p^{kl}\ , \ea where
electric and magnetic charges are in the same $SO(8)$ adjoint
representation as vector fields. Using the definitions in
(\ref{sections}), one obtains\footnote{The expression with explicit
indices is given by \ba\nn\bar P^{ij}_{\ph{ij}kl}=(\tilde
P)^{\ph{kl}ij}_{kl} \ea } \ba\label{} Z_{ij}&=& \frac1{\sqrt2}\left(
(P^{-1/2})^{\ph{ij}kl}_{ij}-(\bar P^{-1/2})^{ij}_{\ph{ij}mn}\bar
y^{mnkl} \right)q_{kl}+ \frac{i}{\sqrt2}\left(
(P^{-1/2})^{\ph{ij}kl}_{ij}+(\bar P^{-1/2})^{ij}_{\ph{ij}mn}\bar
y^{mnkl}
\right)p^{kl}=\nn\\
&=&(P^{-1/2})^{\ph{ij}kl}_{ij} Q_{kl}-(\bar
P^{-1/2})^{ij}_{\ph{ij}mn}
\bar y^{mnkl}\bar Q_{kl}=\nn\\
&=&\frac1{\sqrt2}\left[ \left(\frac1{\sqrt{1-Y\ol
Y}}\right)_{ij}^{\ph{ij}kl}Q_{kl}- \left(\frac1{\sqrt{1-\ol
YY}}\right)^{ij}_{\ph{ij}mn}\bar Y^{mnkl}\bar Q_{kl} \right] \ , \ea
where the complex charges \ba\label{} Q_{ij}\equiv
\frac1{\sqrt2}(q_{ij}+ip^{ij})\ \ea have been introduced.

Then one can also give an expression for the BH
potential, which  is  given by \ba\label{VbhdWN}
V_{BH}&=&\frac12 Z_{ij}\ol Z^{ij}=\nn\\
&=&\frac14\left[
(1-Y\ol Y)^{-1\,ijkl}Q_{kl}\bar Q_{ij}+\right.\nn\\
&&-\left(\sqrt{1-Y\ol Y}\right)^{-1\,ab}_{ij}Q_{ab}\left(\sqrt{1-\ol
Y Y}\right)^{-1\,ij}_{\ph{-1\,ij}cd}Y_{cdkl}
Q_{kl}+\nn\\
&&-\left(\sqrt{1-\ol YY}\right)^{-1\,ij}_{\ph{-1\,ij}ab}\ol Y^{abkl}\bar Q_{kl}\left(\sqrt{1-Y\ol Y}\right)^{-1\ph{ij}cd}_{ij}\bar Q_{cd}+\nn\\
&&+\left.(1-\ol YY)^{-1}_{\ph{-1}ijkl}\ol Y^{ijab}Y_{klmn}\bar
Q_{ab}Q_{mn} \right]\ . \ea Thus, in the expansion around the zero
field configuration, the BH receives contribution from the term
\ba\label{} V_{BH}(\phi=0)&=&\frac14Q_{ij}\bar Q^{ij}\ . \ea The
linear term in the expansion of the BH potential near the point
$\phi=0$ receives contributions from the second and third row of Eq.
(\ref{VbhdWN}), yielding the condition \ba
Q_{ij}\phi_{ijkl}Q_{kl}-\bar Q_{ij}\bar\phi^{ijkl}\bar Q_{kl}&=&0\ ,\\
\dar&&\nn\\
\label{UCLA1}Q_{ij}Q_{kl}\delta_{ijkl}^{mnpq}-\frac1{4!}\bar
Q_{ij}\bar Q_{kl}\e^{ijklmnpq}&=&0\ . \ea The configuration
corresponding to charges $Q_{AB}$ in the singlet of $SU(2)\times
SU(6)$ trivially satisfies condition (\ref{UCLA1}). Furthermore, it
sets to zero the linear term for all values of $\phi$, implying the
$\phi=0$ point to be an attractor point for this configuration.

\section{\label{jazz}$E_{6\left( 6\right) }$-Basis and Relation to $d=5$ }

%\subsection{Kinetic matrix $\mc N$}
This section is aimed to establish the relation between the ${\mc
N}= 8$, $d=4$ theory and ${\mc N}=8$, $d=5$ supergravity
(\cite{Cremmer:1979uq, Sezgin:1981ac}),
especially for what concerns the effective BH potential.

In our normalisations  the kinetic Lagrangian for vector fields in
the ${\mc N}=2$ theory reads (with $\mathcal{F}_{\mu \nu }\equiv
\frac{1}{2}\left( \partial _{\mu }A_{\nu }-\partial _{\nu }A_{\mu
}\right) =\partial _{\lbrack \mu }A_{\nu ]}$)
\cite{CDF-Trieste-Proc,ADF-U-duality-revisited}
\begin{equation}
{\mc L}=\ldots -{\im {\mc N}}_{\Lambda\Sigma}{\mc F}^\Lambda {\mc
F}^\Sigma -{\re {\mc N}}_{\Lambda\Sigma}{\mc F}^\Lambda\ ^\ast {\mc
F}^\Sigma\, ,
\end{equation}
where $\mc N_{\Lambda\,\Sigma}$ is the  $d=4$ vector kinetic matrix,
with $\Lambda, \Sigma=0,1,...,27$. The effective BH potential is
given by \cite{FGK} \ba\label {Vbh} V_{BH}&=&-\frac12Q^{T}\mc M(\mc
N)Q\, , \ea where $Q$ is the symplectic charge vector
$Q=\left(\begin{array}{c}
p^{\Lambda}\\
q_{\Lambda}
\end{array}\right)
$, and the matrix $\mc M$ reads \cite{FGK} \ba\label{} \mc M(\mc
N)&=&\left(
\begin{array}{cc}
&\\
\im\mc N+\re\mc N(\im N)^{-1}\re\mc N \ &\  -\re\mc N(\im\mc N)^{-1}
\\&\\ \vspace{-1pt} -(\im\mc N)^{-1}\re\mc N &
(\im\mc N)^{-1}
\\&\\\
\end{array}\right) .\ea
The $d=5$ $U$-duality group $E_{6(6)}$ acts linearly on the $27$
vectors $\hat A^I_{\hat\mu}$, with $\hat\mu=1,\ldots,5$ and
$I=1,\ldots,27$.
The $d=5$ vector kinetic matrix
$\hat{\mc N}_{IJ}$ is a function of the scalar fields spanning the
$d=5$ scalar manifold $\frac{E_{6\left( 6\right) }}{USp(8)}$ ($dim_{\mathbb{R}}=42,$ \textit{rank}$%
=6$).

According to the splitting $\Lambda=\{0,I\}$, the $d=4$ kinetic vector
matrix assumes the block form \ba\label{} {\mc
N}_{\Lambda\,\Sigma}&=& \left(
\begin{array}{c|c}
&\\
\mc N_{00}\ &\ \mc N_{0\, J}\\&\\
\hline\\ \vspace{-1pt}
\mc N_{I\,0}\ & {\mc N}_{I\, J}\\&\\
\end{array}\right)\ .
\ea
By using to the formul{\ae} obtained in \cite
{Andrianopoli:2002mf} which determine  $\mc N_{\Lambda\,\Sigma}$ in
terms of five-dimensional quantities, in a normalization\footnote{
Compared  to the notation of \cite{Andrianopoli:2002mf}, here we use  ${\mc N}_{\Lambda\Sigma}\to 4{\mc N}_
{\Lambda\Sigma}$,  $2\hat{\mc N}_{IJ}\to a_{IJ}$, $d_{IJK}\to -d_{IJK}/4$ and $a^I\to -a^I$.}
that is suitable for comparison to ${\mc N=2}$ ,  one obtains \ba\label{} {\mc
N}_{\Lambda\,\Sigma}&=& \left(
\begin{array}{c|c}
&\\
\frac13d_{IJK}a^{I}a^{J}a^{K} -i\left( e^{2\phi}{a}_{IJ}a^{I}a^{J}+e^{6\phi}
\right)\ &\
-\frac12d_{IJK}a^{I}a^{K}+i e^{2\phi}{a}_{KJ}a^{K}
\\&\\
\hline\\ \vspace{-1pt}
-\frac12d_{IKL}a^{K}a^{L}+i e^{2\phi}{a}_{IK}a^{K}\ &
\ d_{IJK}a^{K}-
ie^{2\phi}{a}_{IJ}\\&\\
\end{array}
\right)\ . \ea Since the $d_{IJK}$ tensor, the $a^{I}$ fields, the
$d=5$ vector kinetic  matrix $a_{IJ}$ and the field $\phi$
are real, the expressions for $\im\mc N$ and $\re\mc N$ are given by
\ba\label{im-enne} \im\,{\mc N}_{\Lambda\,\Sigma} &=&-
e^{6\phi}\left(
\begin{array}{c|c}
&\\
1+ e^{-4\phi}a_{IJ}\,a^{I}a^{J}
\ &\ -e^{-4\phi}{a}_{KJ}\,a^{K}
\\&\\
\hline\\ \vspace{-1pt}
-e^{-4\phi}a_{IK}\,a^{K}\ &
e^{-4\phi} a_{IJ}\\&\\
\end{array}
\right); \ea
%%%%%%%%%%%%%%%%%%%%%%%%%%%%%%%%%%%%%%%%%%%%%%%%%%%%%%%%%%%%%%%%%%
\ba\label{re-enne} \re\,{\mc N}_{\Lambda\,\Sigma}&=& \left(
\begin{array}{c|c}
&\\
\frac13d_{KLM}a^{K}a^{L}a^{M} \ &\
-\frac12d_{JLM}a^{L}a^{M}\\&\\
\hline\\ \vspace{-1pt}
-\frac12d_{ILM}a^{L}a^{M}\ &
\  d_{IJK}a^{K}\\&\\
\end{array}\right)=\left(
\begin{array}{cc}
\frac13 d & -\frac12 d_{J} \\
-\frac12 d_{I} & d_{IJ}
\end{array}
\right)
\ ,\ea

where the following shorthand notation has been
introduced:\be\label{short} d\equiv d_{IJK}a^I a^J a^K\quad\ ,\quad
d_{I}\equiv  d_{IJK}a^J a^K\quad\ ,\quad d_{IJ}\equiv  d_{IJK}a^{K}\
. \ee The inverse matrix $({\im\,{\mc
N}}_{\Lambda\Sigma})^{-1}\equiv\im\,{\mc N}^{\Lambda\Sigma }$ can be
determined by noticing the block structure of (\ref{im-enne}). Then,
by performing computations analogous to those of
\cite{Ceresole:2007rq}, one finds \ba\label{} (\im{\mc
N}^{-1})^{\Lambda\,\Sigma} &=&-e^{-6\phi} \left(
\begin{array}{c|c}
&\\
1\ &\ a^{J}
\\&\\
\hline\\ \vspace{-1pt} a^{I}\ &
\ a^{I}a^{J}+e^{4\phi}a^{IJ}\\&\\
\end{array}\, ,
\right) \ea where $a^{IJ}\equiv(a_{IJ})^{-1}$.
%%%%%%%%%%%%%%%%%%%%%%%%%%%%%%%%%%%%%%%%%%%%%%%%%%%%%%%%%%%%%%%%%%%%
Inserting the above expressions into Eq. (\ref{Vbh}), the
$\mathcal{N}=8$, $d=4$ effective BH potential can finally be
rewritten in a $d=5$ language:
\ba V_{BH}
\label{poten}
%&=&-\frac12\
%left[ p^{\Lambda}\im\mc N_{\Lambda\Sigma} p^{\Sigma}+
%p^{\Lambda}(\re\mc N\,\im{\mc N^{-1}}\,{\re\mc N}) _{\Lambda\Sigma}p^{\Sigma}+\right.\nn\\
%&&\ph{-\frac12}\left.-2q_{\Lambda}(\im\mc N^{-1}\re\mc N)^{\Lambda}\, _{\Sigma} p^{\Sigma}+q_{\Lambda}
%(\im\mc N^{-1})^{\Lambda\Sigma} q_{\Sigma}\right]=\nn\\
&=&(p^0)^2 \left[  \frac12 e^{2\phi}a_{IJ}a^{I}a^J+ \frac12e^{6\phi}+\frac{1}{8}e^{-6\phi}\left(
\frac{d^2}9+e^{4\phi}a^{IJ}d_I d_J\right)
\right]+\nn\\
&&+p^0p^I\left[ -e^{2\phi}a_{IJ}a^J-\frac14e^{-6\phi} \left(\frac13
dd_I+2e^{4\phi} a^{KJ}d_K d_{JI}\right)
\right]+\nn\\
&&+ p^I p^J\left[ \frac12 e^{2\phi}a_{IJ}+\frac18e^{-6\phi} \left( d_I
d_J+4e^{4\phi} {a}^{KL}d_{IK} d_{LJ}\right)
\right]+\nn\\
&&+\frac16 q_0p^0\, e^{-6\phi}d +\frac16 q_I
p^0\, e^{-6\phi}\left[d\,a^I +3e^{4\phi} a^{KI}d_K \right]+\nn\\
&&-\frac12 q_0p^I\, e^{-6\phi}d_I
- \frac12 q_{I}p^J\, e^{-6\phi}\left[ d_J
a^I+2e^{4\phi} {a}^{KI}d_{JK}
\right]+\nn\\
&&+\frac12(q^0)^2e^{-6\phi}+ q_0q_I e^{-6\phi}a^I+ \frac12 q_I q_J\,
e^{-6\phi}\left[ a^I a^J+e^{4\phi}a^{IJ} \right]\ .
\ea Notice that this formula becomes identical to the corresponding
one of \cite{Ceresole:2007rq} concerning (purely cubic)
$\mathcal{N}=2$ geometries \cite{GST1,GST2}, where
$a_{IJ}=4 e^{4\phi}g_{ij}$ and
${\mc V}\equiv e^{6\phi}$.
%%%%%%%%%%%%%%%%%%%%%%%%%%%%%%%%%%%%%%%%%%%%%%%%%%%%%%%%%%%%

%\subsection{``Axion-free" solutions and relations to BHs and black strings in $d=5$}
\smallskip
The potential (\ref{poten}), because of the definitions (\ref{short}), can be seen to be a polynomial of degree up to sixth in the axion fields, whose general solutions are hard to determine. However, one can consider in particular attractor solutions with vanishing axion fields.
These are given by specific charge configurations that solve
the following attractor equations:
\ba \frac{\p V_{BH}}{\p a^I}\Big|_{a^J=0}&=&
-e^{2\phi}p^0p^K a_{KI}
-e^{-2\phi}q_J p^K d_{IL K} a^{J L}
+q_0q_I e^{-6\phi}=0\ . \ea

Therefore, the BH charge configurations $Q=(p^0,p^I,q_0,q_I)$
supporting axion--free solutions fall into three classes: \ba
a)\quad\quad  &Q_e =&(p^0,0,0,q_I)\quad\quad \textrm{Electric BH}\, ;\nn\\
b)\quad\quad &Q_m =&(0,p^I,q_0,0)\quad\quad \textrm{Magnetic BH}\, ;\nn\\
c)\quad\quad &Q_0 =&(p^0,0,q_0,0)\quad\quad\textrm {KK charged BH}\
. \ea In each of these classes, we now specify the BH potential by setting to zero the appropriate charge configuration in (\ref{poten}):

a) Electric BH:

\be
V_{BH}(\phi,p^0,q_I)\big|_{a^I=0}=
\frac12e^{6\phi}(p^0)^2+\frac12e^{-2\phi}a^{IJ}q_I q_J\, .
\ee

b) Magnetic BH:

\be
V_{BH}(\phi,q_0,p^J)\big|_{a^I=0}=
\frac12e^{-6\phi}(q_0)^2+\frac12 e^{2\phi}a_{IJ}p^I p^J\, .
\ee

c) BH charged with respect to the KK vector:

\be
V_{BH}(\phi,q_0,p^0)\big|_{a^I=0}=
\frac12e^{-6\phi}(q_0)^2+\frac12e^{6\phi}(p^0)^2\ . \ee

In order to recover the complete attractor solution, one also has to
stabilize $e^\phi$.
For the KK charged BH one gets, \be \frac{\p
{V^{KK}_{BH}(\phi,q_0,p^0)}}{\p\phi}\big|_{a^I=0}=0\quad{\iff} \quad
e^{6\phi}=\left|\frac{q_0}{p^0}\right|\ , \ee thus yielding \be
V^{KK}_{BH}(q_0,p^0)\big|_{a^I=0}=|q_0p^0|\ . \ee

In the electric case it holds that
 \be
\frac{\p V_{BH}^{e}}{\p\phi}\big|_{a^I=0}=0\quad\iff \quad
e^{2\phi}=\left( \frac{a^{IJ}q_I q_J}{3(p^0)^2} \right)^{\frac14}\, ,
\ee
implying the critical value
\be \label{Velectr}
V_{BH}^{e}(q_I, p^0)\vert_{a^I=0}=2 \vert p^{0}\vert^{1/2}
\left( \frac{ a^{IJ}q_I q_J}{3}\right)^{3/4}\,  .
\ee
Analogously, for the magnetic BH one finds
\be
\frac{\p{V}_{BH}^{m}}{\p\phi}\big|_{a^I=0}=0 \quad\iff  \quad e^{2\phi}
=\left(
\frac{a_{IJ}p^I p^J}{3q_0^{2}} \right)^{-\frac14}\, ,
\ee
yielding
\be
\label{Vmagn} V_{BH}^{m}(q_0,p^I)\vert_{a^I=0}=
2\vert q_0\vert ^{1/2}
\left( \frac{a_{IJ}p^I p^J}3\right) ^{3/4}\ . \ee

In virtue of the Bekenstein-Hawking entropy-area formula, the above
expressions for the critical electric and magnetic BH potentials
must be compared with appropriate powers of the $E_{6(6)}$ cubic
invariants $\mathcal{I}_3(p)\equiv \frac{1}{3!}d_{IJK}p^Ip^Jp^K$ and
$\mathcal{I}_3(q)\equiv \frac1{3!}d^{IJK}q_Iq_Jq_K$. Indeed, in
$d=5$ it must hold that \cite{AM3}
\be \quad\quad\quad S\sim
V^{3/4}\vert_{crit}\sim |\mathcal{I}_3|^{1/2}\, ,
\ee
Defining the electric and magnetic $d=5$ effective potentials respectively as
\be
V_5^{e}=a^{IJ}q_I q_J\, \quad ,\quad
V_5^{m}=a_{IJ}p^I p^J\, \ee one obtains \be
V^{e}_{crit}=2\vert p^0\vert ^{1/2} \left(
\frac{V_5^{e}}3\right)^{3/4}\big| _{crit} \ee and \be
V^{m}_{crit}=2\vert q^0\vert ^{1/2} \left(
\frac{V_5^{m}}3\right)^{3/4}\big| _{crit}\, . \ee By comparison with
$\mc N=2$ symmetric $d-$geometries having \be
V_5^{e}\vert_{crit}=|\mathcal{I}_3(q)|^{2/3}=|q_1 q_2 q_3|\, , \ee
one obtains the  expressions for the critical potential of the four
dimensional electric and magnetic BHs: \ba\label{}
V^{e}_{BH\,crit}(q_{I},p^{0})&=&
2\sqrt{\frac{|p^{0}d^{IJK}q_{I}q_{J}q_{K}|}{3!}}\ , \ea and \ba\label{}
V^{m}_{BH\,crit}(q_{0},p^{I})=
2\sqrt{\frac{|q_{0}d_{IJK}p^{I}p^{J}p^{K}|}{3!}}\ . \ea

More generally,  these solutions can be compared with the embedding of the $\mc
N=2$ purely cubic supergravities into $\mc N=8$ supergravity, and
using the above critical values of the BH potential in (\ref{entropia}),
one finds for the three family of configurations under exam
the  correct result:
\be\label{SKK}
\frac{S_{BH}}{\pi}=\sqrt{|{\mc I}_4|}\, .
\ee

%\section*{Symplectic sections in d=4}

%From the variation of the 5-dim vielbien given in sec. 3, eq. 3.1b of \cite{Sezgin:1981ac}
%\ba\label{}
%\delta \hat V^{5}_{\mu}&=&-i\epsilon^{a}\gamma^{5}\psi_{\mu a}\ ,
%\ea
%while the variation of vectors reads
%\ba\label{}
%\delta A_{\mu}^{\alpha\beta}&=&\mc V^{\alpha\beta}_{\ph{\alpha\beta}ab}
%(i\eta_{3}\bar\epsilon^{a}\psi^{b}_{\mu})
%\ea

It is interesting to remark that the KK black hole can be connected to the RN
solution by performing an analytic continuation of the charges, as one can  see from the
redefinition  \ba\label{}
p^{0}&\rar&p+ iq\ ,\nn\\
q_{0}&\rar&p-iq\ ,\nn \ea in , which
allows one to recover the RN entropy \ba\label{}
S_{RN}&=\pi(&p^{2}+q^{2})\ . \ea

We conclude this Section by pointing out that the $70$ scalars of $\mathcal{N%
}=8$, $d=4$ supergravity have been decomposed according to representations of $%
USp\left( 8\right) $ (maximal compact subgroup of $E_{6\left(
6\right) }\times SO\left( 1,1\right) $) as follows: \ba &&\quad
\textbf{70}\rar \textbf{42}+\textbf{27}+\textbf{1}\ . \ea The $42$
unstabilized fields are the coordinates of the corresponding moduli
space \cite{Ferrara-Marrani-1}. The non-compact form of the
exceptional group, $E_{6(6)}$, in fact, enters in the expression of
the coset \ba\label{} \frac{E_{6(6)}}{USp(8)}\ , \ea which is the
moduli space of the $d=4$ non-BPS, $Z_{AB}\neq0$ extremal BHs, whose
orbit is precisely \ba\label{} \mc O&=&\frac{E_{7(7)}}{E_{6(6)}}\ .
\ea Indeed, the KK BH is indeed a non supersymmetric solution (see
also Sect. \ref{Intro}).
%We notice that the moduli space of the non-BPS BH with nonzero central charge, as a general
% result, is the scalar manifold of the 5-dim theory, that's why we were able to work with
% the fields in the representation of the 5-dim theory, thus embedding the solution in the higher dimensional $\mc N=8$ Supergravity.

\section{\label{Axion-Dilaton-d=4}Embedding of the Axion-Dilaton Extremal
BH}

The embedding of the axion-dilaton BH in $\mathcal{N}=8$, $d=4$
supergravity can be performed by a three step supersymmetry
reduction, which can be schematically indicated as
\begin{equation}
\mathcal{N}=8\rightarrow \mathcal{N}=4,~n_{V}=6\rightarrow
\mathit{pure}\, \mathcal{N}=4\rightarrow
{\mathcal{N}}=2\quad \mathit{quadratic},~n_{V}=1,  \label{UCB-1}
\end{equation}
where $n_{V}$ denotes the number of vector multiplets coupled to the
supergravity multiplet. More precisely, the first step consists in
truncating $\mathcal{N}=8$ supergravity to an $\mathcal{N}=4$ theory
interacting with six matter (\textit{vector}) multiplets. In the
second step, $\mathcal{N}=4$ reduces to the \textit{pure} theory,
while in the last reduction one obtains $\mathcal{N}=2$ supergravity
\textit{quadratic} \cite{Luciani} theory with a single vector
multiplet.\smallskip

Let us examine more precisely each intermediate step.

1) In the first step, the $\mathcal{N}=8$ central charge matrix
$Z_{AB}$ assumes the block form ($a,b=1,..,4$, $i,j=1,...,4$):
\begin{equation}
Z_{AB}\rightarrow \left(
\begin{array}{ccc}
Z_{ab} &  & 0 \\
&  &  \\
0 &  & i\overline{Z}_{ij}
\end{array}
\right) .  \label{PA-1}
\end{equation}
where $Z_{ab}$ is the $\mathcal{N}=4$ central charge matrix and
$Z_{ij}$ are the matter charges of the $6$ vector multiplets
(sitting in the two-fold antisymmetric of $SU\left( 4\right) $, or
equivalently in the vector representation of $SO\left( 6\right) \sim
SU\left( 4\right) $).

Consequently, the $\mathcal{N}=8$ scalar manifold $\frac{E_{7\left(
7\right) }}{SU\left( 8\right) }$,
%having two \textit{large} orbits,
reduces to
\begin{equation}
\frac{%
SL\left( 2,\mathbb{R}\right) }{U\left( 1\right) }\times
\frac{SO\left( 6,6\right) }{SO\left( 6\right) \times SO\left(
6\right) }=\frac{%
SL\left( 2,\mathbb{R}\right) }{U\left( 1\right) }\times
\frac{SO\left( 6,6\right) }{SU\left( 4\right) \times SU\left(
4\right) }, \label{UCLA!!} \end{equation}
which admits three orbits. This is the scalar manifold
for $\mathcal{N}=4$ supergravity coupled to $6$ vector multiplets,
%which admits three \textit{\ large} charge orbits)
.\smallskip

2) In the second step, the $6$ vector multiplets are eliminated  and $Z_{ij}=0$%
; this corresponds to retaining only states which are singlets with respect to the second $SU(4)$ in the stabilizer of the coset (\ref{UCLA!!})), and the theory becomes pure $=4$, with $U$-duality $SL\left( 2,\mathbb{R}%
\right) \times SU\left( 4\right) $:
\begin{equation}
\left(
\begin{array}{ccc}
Z_{ab}\epsilon  &  & 0 \\
&  &  \\
0 &  & i\overline{Z}_{ij}\epsilon
\end{array}
\right) \rightarrow \left(
\begin{array}{ccc}
Z_{ab} &  & 0 \\
&  &  \\
0 &  & 0
\end{array}
\right) ,  \label{PA-2}
\end{equation}
with $\epsilon =\left( _{-1}^{0}\ _{0}^{1}\right) $. Accordingly,
the scalar
manifold reduces to $\frac{SL\left( 2,\mathbb{R}\right) }{U\left( 1\right) }$%
%, exhibiting only one \textit{large} charge orbit
. Notice that,
the
presence of the axion-dilaton $s$ spanning $\frac{SL\left( 2,\mathbb{R}%
\right) }{U\left( 1\right) }$, in the $\mathcal{N}=4$ supergravity
multiplet, only an  $SU(4)$ out of the  whole (local) $\mathcal{N}=4$
$\mathcal{R}$-symmetry $U(4)
$ gets promoted to (global) $U$-duality symmetry
.\smallskip

3) In the last step, $4$ out of $6$ graviphotons drop out, reducing
the
overall gauge symmetry from $U(1)^{6}$ to $U(1)^{2}$, with resulting $U$%
-duality $SL\left( 2,\mathbb{R}\right) \times U\left( 1\right) $.
Thus, the framework becomes $\mathcal{N}=2$-supersymmetric, with the
two skew-eigenvalues $(Z_{1},Z_{2})$ of $Z_{ab}$ related to the
$\mathcal{N}=2$ central and matter charges $(Z,D_{s}Z)$ :
\begin{equation}
Z_{ab}\rightarrow \left(
\begin{array}{ccc}
Z &  & 0 \\
&  &  \\
0 &  & i\ {\overline{D}}_{\bar{s}}{\bar{Z}}
\end{array}
\right) .  \label{PA-3}
\end{equation}
Therefore, at the $\mathcal{N}=2$ level one can have both BPS attractors ($%
D_{s}Z=0$) and the non-BPS ($Z=0$) ones
\cite{Bellucci:2006xz}.\smallskip\

On a group theoretical side, this step correspond to performing the
decomposition
\begin{equation}
\begin{array}{l}
SU\left( 4\right) \rightarrow SU\left( 2\right) \times SU\left(
2\right)
\times U\left( 1\right) , \\
\\
\mathbf{4}\rightarrow \left(
\mathbf{2},\mathbf{1},\frac{1}{2}\right)
+\left( \mathbf{1},\mathbf{2},-\frac{1}{2}\right) , \\
\\
\mathbf{6}\rightarrow \left( \mathbf{2},\mathbf{2},0\right) +\left( \mathbf{1%
},\mathbf{1},1\right) +\left( \mathbf{1},\mathbf{1},-1\right) ,
\end{array}\label{blues}
\end{equation}
and to retaining only the singlets of $SU\left( 2\right) \times
SU\left( 2\right) $.

%\subsection{\label{large-orbits-moduli-spaces}\textit{Charge
%orbits and \textit{moduli spaces}}
\smallskip
The above three step reduction can be viewed from the point of view
of the classification of  \textit{large} charge orbits
\cite
{Andrianopoli:2006ub,Bellucci:2007ds}.
One starts with the $\mc N=8$ scalar manifold $E_{7(7)}/SU(8)$ admitting the two regular orbits (\ref{O-1/8-BPS-large}) and (\ref{O-non-BPS}).
The \textit{large} charge orbits of $%
\mathcal{N}=4$, $d=4$ supergravity coupled to $6$ vector multiplets
are:
\begin{equation}
\left\{
\begin{array}{l}
\mathcal{O}_{1/4\, BPS}:\quad \quad SL\left( 2,\mathbb{R}\right) \times \frac{%
SO\left( 6,6\right) }{SO\left( 2\right) \times SO\left( 6,4\right) }; \\
\\
\mathcal{O}_{non\, BPS,\ Z_{ab}=0}:\quad \quad SL\left(
2,\mathbb{R}\right) \times \frac{SO\left( 6,6\right) }{SO\left(
2\right) \times SO\left(
6,4\right) }; \\
\\
\mathcal{O}_{non\, BPS,\ Z_{ab}\neq 0}:\quad \quad SL\left(
2,\mathbb{R}\right) \times \frac{SO\left( 6,6\right) }{SO\left(
1,1\right) \times SO\left( 5,5\right) },
\end{array}
\right.  \label{PA-4}
\end{equation}
where the coincidence of the first two orbits is due to the symmetry
between
the gravity and the matter sector.

The corresponding moduli spaces for the $\mathcal{N}=4$, $n=6$
attractor solutions, exploiting the hidden symmetries of the above
charge orbits, are given by:
\begin{equation}
\left\{
\begin{array}{l}
\mathcal{M}_{BPS}=\frac{SO\left( 6,4\right) }{SU(4)\times
SU(2)\times SU(2)};
\\
\\
\mathcal{M}_{non\, BPS, Z_{ab}=0}=\frac{SO\left( 6,4\right) }{SO\left(
6\right)
\times SO\left( 4\right) }; \\
\\
\mathcal{M}_{non\, BPS, Z_{ab}\neq 0}=SO\left( 1,1\right) \times
\frac{SO\left( 5,5\right) }{SO\left( 5\right) \times SO\left(
5\right) }=SO(1,1)\times \frac{SO(5,5)}{USp(4)\times USp(4)}.
\end{array}
\right.  \label{PA-5}
\end{equation}
Notice that $\mathcal{M}_{1/4\, BPS}$ (and $\mathcal{M}%
_{non-BPS,Z_{ab}=0}$) are homogeneous symmetric quaternionic
manifolds, as in the $\mathcal{N}=4\rightarrow \mathcal{N}=2$
reduction they become the hypermultiplets' scalar manifold
\cite{Andrianopoli:2006ub}.

The truncation of the $\mathcal{N}=8$ theory into $\mathcal{N}=4$ is
based on the decomposition
\begin{equation}
E_{7(7)}\rightarrow SL(2,R)\times SO(6,6)
\end{equation}
and on the following group embeddings
\begin{eqnarray}
SO(6,4)\times SO\left( 2\right) &\subsetneq &E_{6\left( 2\right) }; \\
SO(5,5)\times SO\left( 1,1\right) &\subsetneq &E_{6\left( 6\right)
}.
\end{eqnarray}
Therefore, one can readily establish that the orbits 1/4 BPS and non BPS, $%
Z_{ab}=0$ descend from the $\mathcal{N}=8$, BPS orbit
$\frac{E_{7\left(
7\right) }}{E_{6\left( 2\right) }}$, whereas the orbit $\mathcal{O}%
_{nonBPS,\ Z_{ab}\neq 0}$ comes from the $\mathcal{N}=8$, non-BPS orbit $%
\frac{E_{7\left( 7\right) }}{E_{6\left( 6\right) }}$.\smallskip

%\subsection{\label{U-invariants-truncations}Truncations of $U$-invariant}

There is also another way to interpret the  three step reduction (\ref{UCB-1}), that is  in terms of
$U$-duality invariant representations. At group level, the embedding
of the axion-dilaton extremal BH into $\mathcal{N}=8$, $d=4$
supergravity is based on the decomposition of $E_{7(7)}\rightarrow
SU(8)$ and
\begin{equation}
\begin{array}{l}
SU\left( 8\right) \rightarrow SU\left( 4\right) \times SU\left(
4\right)
\times U\left( 1\right) , \\
\\
\mathbf{8}\rightarrow \left(
\mathbf{4},\mathbf{1},\frac{1}{2}\right)
+\left( \mathbf{1},\mathbf{4},-\frac{1}{2}\right) , \\
\\
\mathbf{28}\rightarrow \left( \mathbf{4},\mathbf{4},0\right) +\left( \mathbf{%
6},\mathbf{1},1\right) +\left( \mathbf{1},\mathbf{6},-1\right) , \\
\\
\overline{\mathbf{28}}\rightarrow \left( \overline{\mathbf{4}},\overline{%
\mathbf{4}},0\right) +\left( \mathbf{6},\mathbf{1},-1\right) +\left( \mathbf{%
1},\mathbf{6},1\right) ,
\end{array}
\label{SU(8)-->SU(4)xSU(4)xU(1)}
\end{equation}
where $SU\left( 4\right) \times SU\left( 4\right) \times U\left(
1\right) $ is a maximal subgroup of $SU\left( 8\right) $.

Then, the first truncation ($\mathcal{N}=8\rightarrow \mathcal{N}=4,n=6$) consists in setting
\begin{equation}
\left( \mathbf{4},\mathbf{4},0\right) =0=\left( \overline{\mathbf{4}},%
\overline{\mathbf{4}},0\right) ,  \label{PA-fnight-1}
\end{equation}
which gives rise to the decomposition (\ref{PA-1}).

We recall that the quartic invariant of the $U$-duality group $SL\left( 2,%
\mathbb{R}\right) \times SO\left( 6,n\right) $ of $\mathcal{N}=4$,
$d=4$ supergravity coupled to $n$ vector multiplets is
\cite{Andrianopoli:1997pn}
\begin{equation}
\mathcal{I}_{4}=\mathcal{S}_{1}^{2}-\left| \mathcal{S}_{2}\right|
^{2}, \label{I4-N=4-n}
\end{equation}
where the three $SO\left( 6,n\right) $ invariants $\mathcal{S}_{1}$, $%
\mathcal{S}_{2}$ and $\overline{\mathcal{S}_{2}}$ are defined by ($%
a,b=1,...,4$, $I=1,...,n$):
\begin{eqnarray}
\mathcal{S}_{1} &\equiv &\frac{1}{2}Z_{ab}\overline{Z}^{ab}-Z_{I}\overline{Z}%
^{I};  \label{S1} \\
\mathcal{S}_{2} &\equiv &\frac{1}{4}\epsilon ^{abcd}Z_{ab}Z_{cd}-\overline{Z}%
_{I}\overline{Z}^{I}.  \label{S2}
\end{eqnarray}
The case $n=6$ is remarkably symmetric, as the symmetry of the
gravity and matter sector is the same and furthermore, due to the
isomorphism $SU(4)\sim SO(6)$, the $SO(6)$-vector $Z_{I}$ of matter
charges can be equivalently
represented as the $SU(4)$-antisymmetric tensor $i\overline{Z}_{ij}$ ($%
i,j=1,...,4$).
Consequently, for $n=6$ we have
\begin{eqnarray}
\mathcal{S}_{1,n=6} &\equiv &\frac{1}{2}Z_{ab}\overline{Z}^{ab}-\frac{1}{2}%
\overline{Z}_{ij}Z^{ij};  \label{S1-n=6} \\
\mathcal{S}_{2,n=6} &\equiv &\frac{1}{4}\epsilon ^{abcd}Z_{ab}Z_{cd}-\frac{1%
}{4}\epsilon _{ijkl}Z^{ij}Z^{kl}.  \label{S2-n=6}
\end{eqnarray}
Notice that $\mathcal{%
O}_{1/4BPS}$ and $\mathcal{O}_{nonBPS,\ Z_{ab}=0}$ in Eq.
(\ref{PA-4}) correspond to the two disconnected branches of the same
manifold, classified by the sign of the real $SO\left( 6,6\right)
$-invariant \cite {Andrianopoli:2006ub}
Indeed, $\mathcal{S}_{1,n=6}>0$ for $\mathcal{O}_{1/4BPS}$ and $\mathcal{S}%
_{1,n=6}<0$ for $\mathcal{O}_{nonBPS,\ Z_{ab}=0}$.

By a suitable $U(1)\times SU\left( 4\right) \times SU(4)$
transformation, one can reach the \textit{normal frame} for both
gravity sector and matter sector, such that the two matrices
$Z_{ab}$ and $Z_{ij}$ are simultaneously skew-diagonalized,
obtaining
\begin{eqnarray}
Z_{ab} &\longrightarrow &\left(
\begin{array}{cc}
Z_{1} &  \\
& Z_{2}
\end{array}
\right) \otimes \epsilon ;  \label{PA-night-1} \\
Z_{ij} &\longrightarrow &e^{i\theta }\left(
\begin{array}{cc}
Z_{3} &  \\
& Z_{4}
\end{array}
\right) \otimes \epsilon ,  \label{PA-night-2}
\end{eqnarray}
where $Z_{1},Z_{2}\in \mathbb{R}^{+}$, and $Z_{3},Z_{4}\in \mathbb{R}^{+}$, $%
\theta \in \left[ 0,2\pi \right) $. %
%For reasons which will become
%clear later, in the following treatment we will put
%$Z_{1},Z_{2},Z_{3},Z_{4}$ all on equal footing, treating them as
%complex quantities (even if the phases of
%$Z_{1}$ and $Z_{2}$ can actually be set to zero).
Thus, in the \textit{normal frame} one obtains
\begin{eqnarray}
\mathcal{S}_{1,n=6} &\equiv &\left| Z_{1}\right| ^{2}+\left|
Z_{2}\right|
^{2}-\left| Z_{3}\right| ^{2}-\left| Z_{4}\right| ^{2}; \\
\mathcal{S}_{2,n=6} &\equiv &2\left( Z_{1}Z_{2}-\bar{Z_{3}}\bar{Z_{4}}%
\right) ; \\
\mathcal{I}_{4,n=6} &=&\mathcal{S}_{1,n=6}^{2}-\left| \mathcal{S}%
_{2,n=6}\right| ^{2}=  \nn \\
&=&\sum_{i=1}^{4}\left| Z_{i}\right| ^{4}-2\sum_{i<j=1}^{4}\left|
Z_{i}\right| ^{2}\left| Z_{j}\right| ^{2}+4\left(
\prod_{i=1}^{4}Z_{i}+\prod_{i=1}^{4}\overline{Z_{i}}\right) .
\label{I4-N=4-n=6}
\end{eqnarray}
Eq. (\ref{I4-N=4-n=6}) coincides with the expression of the quartic
invariant of $\mathcal{N}=8$, $d=4$ supergravity, as given by
%the first line of Eq. (2.7) of
\cite{Kallosh:1996uy} (see also \cite{Ferrara:1997ci})
%(thus, the above democratic treatment
%of $Z_{1},Z_{2},Z_{3},Z_{4}$ is now understood, because preparatory
%for the embedding in the $\mathcal{N}=8$ theory).
\smallskip
Considering now the second step of the reduction, where one reaches
the pure $\mathcal{N}=4$ theory, one sets $Z_{ij}=0$, or
equivalently $Z_{3}=0=Z_{4}$ in the \textit{normal frame} (that is, retaining only states which are
singlets with respect to the second $SU(4)$ in the stabilizer of the
coset (\ref{UCLA!!})). Notice that, by doing so,
$\mathcal{I}_{4,n=0}$ becomes a perfect square:
\begin{equation}
\mathcal{I}_{4,n=0}=\mathcal{S}_{1,n=0}^{2}-\left| \mathcal{S}%
_{2,n=0}\right| ^{2}=\left( \left| Z_{1}\right| ^{2}-\left|
Z_{2}\right| ^{2}\right) ^{2}=\left( Z_{1}^{2}-Z_{2}^{2}\right)
^{2}. \label{I4-pure-central-charge-basis}
\end{equation}
Eq. (\ref{I4-pure-central-charge-basis}) implies that
$\mathcal{I}_{4,n=0}$
is (weakly) positive, and as a consequence an unique class of  \textit{%
large} attractor exists, namely the $1/4$-BPS one. The
(weak) positivity of $\mathcal{I}_{4,n=0}$ is consistent with
the known expression of $\mathcal{I}_{4,n=0}$ in terms of the
magnetic and electric charges $\left( p^{\Lambda },q_{\Lambda
}\right) $ ($\Lambda =1,...,6$):
\begin{equation}
\mathcal{I}_{4,n=0}=4\left[ p^{2}q^{2}-\left( p\cdot q\right)
^{2}\right] , \label{I4-n=0-bare-charges-basis}
\end{equation}
where here $p^{2}\equiv p^{\Lambda }p^{\Sigma }\delta _{\Lambda \Sigma }$, $%
q^{2}\equiv q_{\Lambda }q_{\Sigma }\delta ^{\Lambda \Sigma }$ and
$p\cdot q\equiv p^{\Lambda }q_{\Sigma }\delta _{\Sigma }^{\Lambda
}$. Notice that in the basis of \textit{bare} charges
$\mathcal{I}_{4,n=0}$, as given by Eq.
(\ref{I4-n=0-bare-charges-basis}), is (weakly) positive due to the
\textit{Schwarz inequality}, and not because it is a non-trivial
perfect square of an expression of the \textit{bare} magnetic and
electric charges \cite{Gnecchi-1}.

Notice that $\sqrt{\mathcal{I}_{4,n=0}}$ (with $\mathcal{I}_{4,n=0}$
given by Eq. (\ref{I4-n=0-bare-charges-basis})) must coincide with the
value of the effective BH potential of the pure $\mathcal{N}=4$
theory at its critical points. This can be understood  (see the
recent discussion given in \cite{Andrianopoli:2006ub} and
\cite{FHM1-Erice-07}) because this potential reads as follows ($\Lambda
=1,...,6$):
\begin{eqnarray}
V_{BH,pure\, \mathcal{N}=4}\left( {\phi ,\,a,\,p^{\Lambda },\,q_{\Lambda }}%
\right)  &=&e^{2\phi }(sp_{\Lambda }-q_{\Lambda })(\bar{s}p^{\Lambda
}-q^{\Lambda })=  \nn\\
&=&(e^{2\phi }a^{2}+e^{-2\phi })p^{2}+e^{2\phi }q^{2}-2ae^{2\phi
}p\cdot q, \label{Thu-14}
\end{eqnarray}
where the complex (axion-dilaton) field
\begin{equation}
s\equiv a+ie^{-2\phi }
\end{equation}
parametrizes the coset $\frac{SU\left( 1,1\right) }{U\left( 1\right) }$ of $%
\mathcal{N}=4$, $d=4$ \textit{pure} supergravity \cite{CSF}.

By computing the criticality conditions of
$V_{BH,pure\, \mathcal{N}=4}$, one obtains the following stabilization
equations for the axion $a$ and the dilaton $\phi $ at criticality, ${( \phi ,a) =\left( {\phi }_{H}( p,q) {,\,a}_{H}( p,q) \right) }$:
\cite{Andrianopoli:2006ub}:
\ba
\frac{    {\partial V}_{BH}( \phi , a, p, q) } {\partial a }\big| _{crit}
   &=& 0\Longleftrightarrow a_{H}( p,q) =
\frac{p\cdot q}{p^2};  \label{Thu-16}\\
\frac{{\partial V}_{BH}( {\phi ,\,a,\,p,\,q})}{{\partial\phi }}\big| _{crit }
&=& -e^{-4\phi }p^2+q^2-a_{H}( p,q) p\cdot q=-e^{-4\phi}p^{2}+q^{2}-\frac{( p\cdot q) ^{2}}{p^{2}}=0;  \nn \\
&\Updownarrow &  \nn \\
e^{-2\phi _{H}\left( p,q\right) }&=&\frac{\sqrt{p^{2}q^{2}-(p\cdot q)^{2}}}{p^2}.  \label{Thu-17}
\ea
\newline
Thus, the Bekenstein-Hawking BH entropy is computed to be
\begin{equation}
S_{BH}\left( p,q\right) =\frac{A_{H}\left( p,q\right) }{4}=\pi
V_{BH}\left(
\phi _{H}\left( p,q\right) ,a_{H}\left( p,q\right) ,p,q\right) =2\pi {\sqrt{%
p^{2}q^{2}-(p\cdot q)^{2}}=\pi }\sqrt{\mathcal{I}_{4,n=0}}.
\end{equation}
\smallskip\

The third and last step, when the pure $\mathcal{N}=4$ theory
reduces to the $\mathcal{N}=2$ quadratic theory with $n_{V}=1$, is
performed through the truncation $\left( U\left( 1\right) \right)
^{6}\rightarrow \left( U\left( 1\right) \right) ^{2}$ of the overall
Abelian gauge invariance ($\Lambda =1,...,6\rightarrow \Lambda
=1,2$). In this case, $\mathcal{I}_{4,n=0,\left( U\left( 1\right)
\right) ^{6}\rightarrow \left(
U\left( 1\right) \right) ^{2}}$ is a perfect square in both the basis of $%
Z_{ab}$ and in the basis of charges $\left( p^{\Lambda },q_{\Lambda
}\right)
$, and it actually is the square of the \textit{quadratic} invariant $%
\mathcal{I}_{2(n=1)}$ of the axion-dilaton system:
\ba \mathcal{I}_{4,n=0,\left( U\left( 1\right) \right) ^{6}\rightarrow \left( U\left( 1\right)
\right) ^{2}}&=&\left( \left| Z_{1}\right| ^{2}-\left|
Z_{2}\right| ^{2}\right) ^{2}=4\left( p^{1}q_{2}-p^{2}q_{1}\right) ^{2}=%
\mathcal{I}_{2(n=1)}^{2}; \\
&\Updownarrow&  \nn\\
\mathcal{I}_{2(n=1)}&=&\pm 2\left| p^{1}q_{2}-p^{2}q_{1}\right| ,
\ea
implying that the axion-dilaton system exhibits two types of
attractors: the
$\frac{1}{2}$-BPS one ($\mathcal{I}_{2(n=1)}>0$) and the non-BPS $Z=0$ one (%
$\mathcal{I}_{2(n=1)}<0$).

By further putting
\be
p^{1}=0=q_{2},~p^{2}\equiv p,~q_{1}\equiv q  \label{PA-nightt-1}
\end{equation}
($\Rightarrow $ $p\cdot q=0$), one obtains:
\ba
{\mathcal {I}}^{\ast}_{4\left(n=0, U( 1) ^{6}\rightarrow U( 1) ^{2}\right)} &=& {\mathcal{I}}^{2\ast}_{2(n=1)}=4\left( pq\right) ^{2}; \\
&\Updownarrow&   \nn \\
\mathcal{I}^{\ast}_{2(n=1)} &=&\pm 2\left| pq\right| ,
\label{PA-nightt}
\ea
where ${\mc I}^{\ast}$ means the
evaluation along Eq. (\ref{PA-nightt-1}). For a recent treatment of
the axion-dilaton-Maxwell-Einstein-(super)gravity system and of the
extremal BH attractors therein, see \textit{e.g.} Sects. 6 and 7 of
\cite{FHM1-Erice-07}.

The similarity between the r.h.s.'s of Eqs. (\ref {I4-KK}) and
(\ref{PA-nightt}) is only apparent. In fact, the KK extremal BH has
$\sqrt{-\mathcal{I}_{4,KK}}$, which necessarily implies that it is
non-BPS ($Z_{AB}\neq 0$ in $\mathcal{N}=8$ and $Z\neq 0$ in
$\mathcal{N}=2$). On the other hand, the axion-dilaton extremal BH
has $\mathcal{I}^\ast_{2(n=1)} $ and a \textit{``}$\pm $\textit{''}
in the r.h.s.,  so that it can be both $\frac{1}{2}$-BPS and non-BPS
$Z=0$ in $\mathcal{N}=2$. Moreover, the choice (\ref{PA-nightt-1})
leads to vanishing axion $a$ (see Eq. (\ref{Thu-16})), and this
explains that  Eqs. (\ref{PA-nightt}) has $SO\left( 1,1\right)$
symmetry, as Eq. (\ref{I4-KK}).

\subsection{\label{scalar-truncations}Truncations of the scalar sector}

As reported \textit{e.g.} in Sects. 6 and 7 of \cite
{FHM1-Erice-07}, one can see that the attractor mechanism stabilizes the
complex axion-dilaton $s$ at the event-horizon of the axion-dilaton
extremal BH itself, while, as given by Eqs. (\ref{r^3_H-KK}) and
(\ref{a_H-KK}) within the branching (\ref{SU(8)-->USp(8)-70}), only
one real scalar degree of freedom, namely the KK radius $r_{KK}$
defined by Eq. (\ref{r-KK}), is stabilized at the event horizon of
the extremal KK BH.

The relevant branching of the scalar sector
for the embedding of the axion-dilaton extremal BH into $\mathcal{N}=8$, $%
d=4 $ supergravity is given by:
\begin{equation}
\begin{array}{l}
SU(8)\rightarrow SU\left( 4\right) \times SU\left( 4\right) \times
U\left(
1\right) , \\
\\
\mathbf{70}\rightarrow \left( \mathbf{1,1,}2\right) +\left( \mathbf{1,1,-}%
2\right) +\left( \mathbf{6,6,}0\right) +\left( \overline{\mathbf{4}}\mathbf{%
,4,}1\right) +\left(
\mathbf{4,}\overline{\mathbf{4}}\mathbf{,}-1\right) .
\end{array}
\label{SU(8)-->SU(4)xSU(4)xU(1)-70}
\end{equation}
Eq. (\ref{SU(8)-->SU(4)xSU(4)xU(1)-70}) is the analogue of Eqs.
(\ref {SU(8)-->SU(6)xSU(2)xU(1)-70}) and (\ref{SU(8)-->USp(8)-70}),
holding respectively for the ($\mathcal{N}=8$, $d=4$ embedding of
the) RN and KK $d=4$ extremal (and asymptotically flat) BHs.

A remarkable feature characterizing the branchings (\ref
{SU(8)-->SU(6)xSU(2)xU(1)-70}), (\ref{SU(8)-->USp(8)-70}) and (\ref
{SU(8)-->SU(4)xSU(4)xU(1)-70}) is the possible presence of a singlet
in their r.h.s.'s. The decomposition
(\ref{SU(8)-->SU(4)xSU(4)xU(1)-70}) contains two $SU\left( 4\right)
\left( \times SU\left( 4\right) \right) $ singlets, whereas the
decomposition (\ref{SU(8)-->USp(8)-70}) contains a real singlet, and
the decomposition (\ref{SU(8)-->SU(6)xSU(2)xU(1)-70}) does not
contain any singlet.  The presence of the
singlet may lead to an underlying maximal
compact symmetry ($U\left( 1\right) $ for
(\ref{SU(8)-->SU(6)xSU(2)xU(1)-70}), absent for
(\ref{SU(8)-->USp(8)-70}), and $SU\left( 4\right) $ for (\ref
{SU(8)-->SU(4)xSU(4)xU(1)-70})).

\begin{enumerate}
\item  The first truncation ($\mathcal{N}=8\rightarrow \mathcal{N}=4,n_V=6$)
corresponds to setting\footnote{%
Notice the difference with respect to the analogue truncation
condition (\ref
{PA-fnight-1}) for the decomposition of the $\mathbf{28}$ and $\overline{%
\mathbf{28}}$ of $SU\left( 8\right) $.}
\begin{equation}
\left( \overline{\mathbf{4}}\mathbf{,4,}1\right) =0=\left( \mathbf{4,}%
\overline{\mathbf{4}}\mathbf{,}-1\right) .  \label{PA-fnight-2}
\end{equation}
Indeed, by applying the condition (\ref{PA-fnight-2}), one obtains
the
correct quantum numbers of the scalar manifold $\frac{SL\left( 2,\mathbb{R}%
\right) }{U\left( 1\right) }\times \frac{SO\left( 6,6\right)
}{SO\left( 6\right) \times SO\left( 6\right) }$ of the
$\mathcal{N}=4$, $d=4$ supergravity coupled to $6$ vector
multiplets.

\item  The second truncation ($\mathcal{N}=4,n_V=6\rightarrow \textit{pure} \mathcal{N}=4$
) simply consists in implementing the condition
\begin{equation}
\left( \mathbf{6,6,}0\right) =0,  \label{PA-fnight-3}
\end{equation}
which is consistently symmetric under the exchange of the
gravity sector and the matter sector. Through condition
(\ref{PA-fnight-3}), one achieves the correct quantum numbers of the
scalar manifold $\frac{SL\left(
2,\mathbb{R}\right) }{U\left( 1\right) }$ of the \textit{pure} $\mathcal{N}%
=4 $, $d=4$ supergravity.

\item  The third and last step (\textit{pure} $\mathcal{N}=4\rightarrow
\mathcal{N}=2$\textit{quadratic}, $n_{V}=1$) does not change anything
with respect to the previous one. Indeed, the scalar sector is
unaffected by this
third truncation, and the scalar manifold remains $\frac{SL\left( 2,\mathbb{R%
}\right) }{U\left( 1\right) }$. \textbf{\ }
\end{enumerate}

\section{\label{Conclusion}Conclusions}
In the present investigation, we have considered some examples of
extremal BH configurations in the framework of BH attractors of
$\mathcal{N}=8$ supergravity.

The effective BH potential has been computed in different bases,
namely in
the manifestly $SU\left( 8\right) $-coveriant basis, as well as in the $%
USp\left( 8\right) $-covariant one. The former is suitable to
describe the (BPS) Reissner-N\"{o}rdstrom extremal BH with its
$U\left( 1\right) $
symmetry, as a consequence of the attractor point to be the origin of the $%
d=4$ scalar manifold $\frac{E_{7\left( 7\right) }}{SU\left( 8\right)
}$. The latter has $d=5$ origin, and it is appropriate in order to
describe the non-BPS Kaluza-Klein  extremal BH, with its
$SO\left( 1,1\right) $ symmetry arising from the non-trivial
attractor value of the KK radial mode.

We have also considered the axion-dilaton system, whose BPS or
non-BPS
nature depends on whether it is embedded in $\mathcal{N}=2$ \textit{%
quadratic} or in $\mathcal{N}=4$, $d=4$ supergravity. The
axion-dilaton extremal BH is obtained as a particular case of the
attractor equations of the maximal $d=4 $ theory.  In that case, all $70$
scalars other than the $SU\left( 4\right) \times SU\left( 4\right)
$-singlets in the decomposition \ref {SU(8)-->SU(4)xSU(4)xU(1)-70}
are set to vanish, and correspondingly only $12 $ graviphoton
electric and magnetic charges are taken  to be nonzero  (see Eq. (\ref
{SU(8)-->SU(4)xSU(4)xU(1)})). At the level $\mathcal{N}=2$,  this
attractor solution is obtained by retaining only $4$ ($2$ electric
and $2$ magnetic) non-vanishing charges, according to the decomposition (\ref{blues}%
) of $SU\left( 4\right) $.

In Appendix A, we have finally considered the embedding of the $stu$ model in $\mathcal{N}=8$, $%
d=4$ and $d=5$ supergravity, is considered. In the $d=4$ case, all
non-singlet charges in the decomposition of $E_{7\left( 7\right) }$
with respect to $SO\left( 4,4\right) \times \left( SL\left(
2,\mathbb{R}\right) \right) ^{3}$ are set to vanish \cite{ADFFT},
whereas for $d=5$ one obtains
an axion-free framework, given by non- zero values for $(p^{0}$,$)q_{1}$,$q_{2}$,$%
q_{3}$ or $(q_{0}$,$)p^{1}$,$p^{2}$,$p^{3}$ .

\section*{Acknowledgments}

This work is supported in part by the ERC Advanced Grant no. 226455, \textit{%
``Supersymmetry, Quantum Gravity and Gauge Fields''}
(\textit{SUPERFIELDS}).

A. M. wishes to thank R. Kallosh for enlightening discussions. A. M.
would like to thank the \textit{William I. Fine Theoretical Physics
Institute} (FTPI) of the University of Minnesota, Minneapolis, MN
USA, the \textit{Center for Theoretical Physics} (CTP) of the
University of California, Berkeley, CA USA, the Department of
Physics, \textit{Theory Unit Group} at CERN, Geneva CH, and the
Department of Physics and Astronomy of the University of California,
Los Angeles, CA USA, where part of this work was done, for kind
hospitality and stimulating environment.

The work of A. C. ~has been supported in part by MIUR-PRIN contract
20075ATT78, while the work of S. F.~has been supported in part by D.O.E.~grant
DE-FG03-91ER40662, Task C. Finally,  the work of A. M. has been supported by an INFN visiting Theoretical
Fellowship at SITP, Stanford University, Stanford, CA USA.

%\ba
%\re\mc N&=&
%\left(
%\begin{array}{cc}
%d
%\end{array}
%\right)
%\ea

%\begin{appendix}
\appendix \setcounter{equation}0

\section{\label{stu5}Appendix\\Truncation of $\mathcal{N}=8,d=5$ supergravity\\to the $%
d=5$ uplift of the $stu$ model}

The bosonic sector of the $\mathcal{N}=8$, $d=5$ supergravity theory
consists in the metric $g_{\mu \nu }$ ($\mu ,\nu =1,...,5$), $27$ vectors $%
A_{\mu }^{\Lambda }$ and $42$ scalars $\phi _{abcd}$ parametrizing
the coset
$\frac{E_{6(6)}}{USp(8)}$. The index $\Lambda =1,\ldots 27$ is in the $%
\mathbf{27}$ of $E_{6(6)}$, and it can be traded for a couple of
flat antisymmetric indices $(ab)$ of $USp(8)$. Thus, the vectors
$A_{\mu }^{ab}$ transform in the $\mathbf{27}$ of $USp(8)$ , that is
\begin{equation}
\mathbf{27~}\mathrm{of~}E_{6(6)}\longrightarrow \mathbf{27~}\mathrm{of~}%
USp(8)\,.
\end{equation}
The $42$ scalars $\phi _{abcd}$ are in the traceless self-real
$4$-fold antisymmetric representation $\mathbf{42}$ of $USp(8)$.

Upon performing the $d=5\rightarrow d=4$ reduction, one gets $70$
scalars, which split into the following irreps. of $USp(8)$:
\begin{equation}
\mathbf{70}=\mathbf{42}+\mathbf{27}+\mathbf{1}\,.
\end{equation}
Here $\mathbf{27}$ accounts for the \textit{axions} coming from the $%
A_{5}^{ab}$ vectors of $E_{6(6)}$, ${\ }\mathbf{1}$ is the KK radius
$r_{KK}$ (see the definition (\ref{r-KK})), and $\mathbf{42}$
corresponds to the scalars in $\frac{E_{6(6)}}{USp(8)}$.

In order to extract the $stu$ model, we notice that its $d=5$ uplift
is the $\left( SO(1,1)\right) ^{2}$ model with cubic hypersurface
\cite{GST1,GST2} (see \textit{e.g.} the treatment given in
\cite{Ceresole:2007rq})
\begin{equation}
\widehat{\lambda }^{1}\widehat{\lambda }^{2}\widehat{\lambda }^{3}=1.\label%
{stu-d=5}
\end{equation}

The $\mathcal{N}=8\longrightarrow \mathcal{N}=2$, $d=5$
supersymmetry reduction corresponds, at the level of $E_{6(6)}$, to
taking the decomposition
\begin{equation}
E_{6(6)}\longrightarrow SO(1,1)\times SO(5,5)\longrightarrow \left(
SO(1,1)\right) ^{2}\times SO(4,4)\,,\label{UCLA!-1}
\end{equation}
so that (weights with respect to $SO(1,1)$'s are disregarded,
irrelevant for our purposes)
\begin{equation}
\mathbf{27}\rightarrow \mathbf{1}+\mathbf{16}+\mathbf{10}\rightarrow \mathbf{%
1}+\mathbf{8}_{s}+\mathbf{8}_{c}+\mathbf{1}+\mathbf{1}+\mathbf{8}_{v}.\label%
{UCLA!-2}
\end{equation}

Thus, three $SO(4,4)$-singlets are generated; they correspond to the
three Abelian vector fields of the $d=5$ uplift of the $stu$ model.
By further
reducing to $d=4$, one gets a further vector from the KK vector (\textit{%
alias} the $d=4$ graviphoton). This can be easily seen by completing
the
decomposition (\ref{UCLA!-1}) starting from the $U$-duality group $%
E_{7\left( 7\right) }$ of $d=4$ maximal supergravity:
\begin{equation}
E_{7(7)}\longrightarrow SO\left( 1,1\right) \times
E_{6(6)}\longrightarrow \left( SO(1,1)\right) ^{2}\times
SO(5,5)\longrightarrow \left( SO(1,1)\right) ^{3}\times SO(4,4)\,,
\end{equation}
so that Eq. (\ref{UCLA!-2}) gets completed as (as above, neglecting
weights with respect to $SO(1,1)$, as they are irrelevant for our
purposes)
\begin{equation}
\mathbf{28}\rightarrow \mathbf{27}+\mathbf{1}\rightarrow \mathbf{1}+\mathbf{%
16}+\mathbf{10}+\mathbf{1}\rightarrow \mathbf{1}+\mathbf{8}_{s}+\mathbf{8}%
_{c}+\mathbf{1}+\mathbf{1}+\mathbf{8}_{v}+\mathbf{1},\label{UCLA!-3}
\end{equation}
containing four $SO(4,4)$ singlets in the last term.

It is worth pointing out that at $d=4$ the $\left( SO\left(
1,1\right) \right) ^{3}$ commuting with $SO\left( 4,4\right) $ gets enhanced to $\left( SL\left( 2,\mathbb{R}%
\right) \right) ^{3}$. By further decomposing
\begin{equation}
SO\left( 4,4\right) \rightarrow \left( SL\left( 2,\mathbb{R}\right)
\right) ^{4},
\end{equation}
this yields the $\left( SL\left( 2,\mathbb{R}\right) \right) ^{7}$,
used for the seven qubit entanglement in quantum information theory
\cite {Ferrara-Duff-QIT,Levay-QIT}.

Notice that the presence of three different $\mathbf{8}$'s of
$SO(4,4)$ in
the r.h.s. of the decomposition (\ref{UCLA!-2}) (as well as of (\ref{UCLA!-3}%
)) is the origin of the \textit{triality }symmetry \cite{DLR,BKRSW}
of the $stu$ model \cite{ADFFT}.

The $\left( SO(1,1)\right) ^{2}$ factor in the r.h.s. of the
branching (\ref {UCLA!-1}) is nothing but the scalar manifold of the
$d=5$ counterpart of the $stu$ model (spanned by $\widehat{\lambda }^{1}$, $\widehat{\lambda }^{2}$ and $%
\widehat{\lambda }^{3}$ satisfying the cubic constraint
(\ref{stu-d=5})). On the other hand, the $\left( SO(1,1)\right)
^{3}$ factor in the r.h.s. of the branching (\ref{UCLA!-3}) is
spanned by the (unconstrained, strictly
positive) $d=4$ \textit{dilatons} $\lambda ^{1}\equiv -Im\left( s\right) $, $%
\lambda ^{2}\equiv -Im\left( t\right) $ and $\lambda ^{3}\equiv
-Im\left( u\right) $. They are related to their hatted counterparts
by $\lambda ^{i}\equiv r_{KK}\widehat{\lambda }^{i}$, $i=1,2,3$,
implying (see Eqs. (\ref {stu-d=5}) and Eq. (\ref{r-KK}); see also
\textit{e.g.} \cite {Ceresole:2007rq}))
\begin{equation}
\lambda ^{1}\lambda ^{2}\lambda ^{3}=r_{KK}^{3}\equiv \mathcal{V}.\label%
{UCLA-cubic}
\end{equation}

The decomposition of the $d=5$ stabilizer (analogue to the decomposition (%
\ref{UCLA!-1}) of the $U$-duality group of the $d=5$ maximal
supergravity) reads as follows:
\begin{equation}
USp(8)\rightarrow USp(4)\times USp(4)=Spin(5)\times
Spin(5)\rightarrow Spin(4)\times Spin(4)=\left( SU(2)\right)
^{2}\times \left( SU(2)\right) ^{2},
\end{equation}
yielding the following decomposition of the fundamental $\mathbf{8}$ of $%
USp(8)$:
\begin{equation}
\mathbf{8}\rightarrow (\mathbf{4},\mathbf{1})+(\mathbf{1},\mathbf{4}%
)\rightarrow (\mathbf{2},\mathbf{1},\mathbf{1},\mathbf{1})+(\mathbf{1},%
\mathbf{2},\mathbf{1},\mathbf{1})+(\mathbf{1},\mathbf{1},\mathbf{2},\mathbf{1%
})+(\mathbf{1},\mathbf{1},\mathbf{1},\mathbf{2})\,.
\end{equation}
This allows one to compute the corresponding branchings of the $\mathbf{27}%
=\left( \mathbf{8}\times \mathbf{8}\right) _{A,0}$ and
$\mathbf{42}=\left( \mathbf{8}\times \mathbf{8}\times
\mathbf{8}\times \mathbf{8}\right) _{A,0}$
(the subscript ``$A,0$'' standing for ``antisymmetric traceless'') of $%
USp\left( 8\right) $ (the intermediate decompositions with respect to $%
USp(4)\times USp(4)$ are omitted, because irrelevant for our
purposes):
\begin{eqnarray}
\mathbf{27} &\rightarrow &(\mathbf{2},\mathbf{2},\mathbf{1},\mathbf{1})+(%
\mathbf{2},\mathbf{1},\mathbf{2},\mathbf{1})+(\mathbf{2},\mathbf{1},\mathbf{1%
},\mathbf{2})+(\mathbf{1},\mathbf{2},\mathbf{2},\mathbf{1})+  \nn \\
&&+(\mathbf{1},\mathbf{2},\mathbf{1},\mathbf{2})+(\mathbf{1},\mathbf{1},%
\mathbf{2},\mathbf{2})\,+3\,(\mathbf{1},\mathbf{1},\mathbf{1},\mathbf{1});%
\label{UCLA-27} \\
&&  \nn \\
\mathbf{42} &\rightarrow &(\mathbf{2},\mathbf{2},\mathbf{2},\mathbf{2})+(%
\mathbf{2},\mathbf{2},\mathbf{1},\mathbf{1})+(\mathbf{2},\mathbf{1},\mathbf{2%
},\mathbf{1})+(\mathbf{2},\mathbf{1},\mathbf{1},\mathbf{2})+(\mathbf{1},%
\mathbf{2},\mathbf{2},\mathbf{1})+  \nn \\
&&+(\mathbf{1},\mathbf{2},\mathbf{1},\mathbf{2})+(\mathbf{1},\mathbf{1},%
\mathbf{2},\mathbf{2})\,+2\,(\mathbf{1},\mathbf{1},\mathbf{1},\mathbf{1})%
\label{UCLA-42}
\end{eqnarray}
Consistently with previous statements, the three $\left( SU(2)\right) ^{4}$%
-singlets in the r.h.s. of the decomposition (\ref{UCLA-27}) and the two $%
\left( SU(2)\right) ^{4}$-singlets in in the r.h.s. of the decomposition (%
\ref{UCLA-42}) respectively are the three Abelian vector fields
(including
the $d=5$ graviphoton) and the two independent real scalars (say, $\widehat{%
\lambda }^{1}$ and $\widehat{\lambda }^{2}$) in the bosonic spectrum of the $%
\left( SO\left( 1,1\right) \right) ^{2}$ model, which is the $d=5$
uplift of the $stu$ model.

Reducing to $d=4$, the six real scalar degrees of freedom of the
$stu$ model are the radius $r_{KK}$ (see Eqs. (\ref{r-KK}) and
(\ref{UCLA-cubic})), the two scalars $\widehat{\lambda }^{1}$ and
$\widehat{\lambda }^{2}$, and the
three \textit{axions} (coming from the fifth component $A_{5}^{I}$ ($I=1,2,3$%
) of the three $d=5$ vectors). As previously mentioned, the four
$d=4$ vectors come from the three $d=5$ vectors and from the KK
vector $g_{5\mu }$ ($\mu =1,...,4$).

Finally, it should be notice that $\lambda ^{1}\lambda ^{2}\lambda
^{3}$
(defining the volume of the $d=5$ cubic hypersurface through Eqs. (\ref{r-KK}%
) and (\ref{UCLA-cubic})) can be obtained through a consistent
truncation of the $E_{6(6)}$-invariant expression ($\Lambda ,\Sigma
,\Delta =1,...,27$)
\begin{equation}
\frac{1}{3!}d_{\Lambda \Sigma \Delta }\lambda ^{\Lambda }\lambda
^{\Sigma }\lambda ^{\Delta }
\end{equation}
to $\left( SO\left( 1,1\right) \right) ^{2}$, by retaining only the
three singlets of $SO\left( 4,4\right) $ (see the decompositions
(\ref{UCLA!-1}) and (\ref{UCLA!-2}) above).
%\end{appendix}

\end{document}